\documentclass[onecolumn]{aa}

\usepackage{graphicx}
\usepackage{txfonts}
\usepackage{hyperref}
\usepackage{natbib}
\usepackage{soul}
\usepackage{xcolor}

\newcommand{\elsasser}{Els\"asser}
\newcommand{\alfven}{Alfv\'en}

\begin{document}

   \title{Uniturbulence and \alfven{} wave solar model}
   \titlerunning{UAWSOM}


   \author{T. Van Doorsselaere
          \inst{1}
          \and
          M. V. Sieyra\inst{2}
          \and
          N. Magyar\inst{1}
          \and
          M. Goossens \inst{1}
          \and 
          Luka Banovi\'c \inst{1}
          }

   \institute{Centre for mathematical Plasma Astrophysics, Department of Mathematics, KU~Leuven, Celestijnenlaan 200B, B-3001 Leuven, Belgium\\
              \email{tom.vandoorsselaere@kuleuven.be}
         \and
             D\'epartement d'Astrophysique/AIM, CEA/IRFU, CNRS/INSU, Universit\'e Paris-Saclay, Universit\'e de Paris, F-91191,
Gif-sur-Yvette, France
             }

   \date{Received ; accepted }

 
  \abstract
   {AWSOM-type models \citep{vanderholst2014} have been very successful in describing the solar atmosphere by incorporating the \alfven{} wave 
driving as extra contributions in
the global MHD equations. However, they lack the contributions from other wave modes.}
   {In this paper, we aim to write governing equations for the energy evolution equation of kink waves. In a similar
manner as AWSOM, we combine the kink wave evolution equation with MHD. Our goal is to incorporate the extra heating provided by the
uniturbulent damping of the kink waves. We attempt to construct the UAWSOM equations (uniturbulence and \alfven{} wave driven solar models).}
   {We have recently described the MHD equations in terms of the $Q$-variables. These latter variables allow to follow the evolution of waves in 
a co-propagating reference frame. We
transform the $Q$-variable MHD equations into an energy evolution equation. First we do this generally, and then we specialise to the description of
kink waves. We model the resulting UAWSOM system of differential equations in a 1D solar atmosphere configuration in a python code. We also couple 
this evolution equation to the slowly varying MHD formulation and solve the system in 1D.}
   {We find that the kink wave energy evolution equation contains non-linear terms, even in the absence of counterpropagating waves. Thus, we confirm
earlier analytical and numerical results. The non-linear damping is expressed solely through equilibrium parameters, rather than an ad-hoc 
perpendicular correlation term (popularly quantified with a length scale $L_\perp$), as in the case of the AWSOM models.
We have combined the kink evolution equation with the MHD equations to obtain the UAWSOM equations. A proof-of-concept numerical implementation in 
python shows
that the kink wave driving indeed leads to radial outflow and heating. Thus, UAWSOM may have the necessary ingredients to drive the solar wind and 
heat the
solar corona against losses.}
   {Not only does our current work constitute a pathway to fix shortcomings in heating and wind driving in the popular AWSOM model, it also provides
the mathematical formalism to incorporate more wave modes (e.g. the parametric decay instability) for additional driving of the solar wind.}


   \maketitle
%

\section{Introduction} \label{sec:intro}
There are observational indications that the solar corona and the solar wind are filled with \alfven{} waves. In-situ observations allow us 
to measure \alfven{} waves directly in the solar wind \citep{bruno2013}. However, in the corona they are usually only 
identified as spectral line broadenings. From spectroscopic measurements, it was thought that the \alfven{} waves' energy is considerable
\citep{banerjee1998,hahn2013,pant2020}. Only recently, a few direct measurements have become available \citep{depontieu2012,shetye2021,petrova2024}. 
As such, the true energy content of \alfven{} waves in the corona is not so well established. 

On the other hand, high-resolution imaging observations from SDO/AIA and SolO/EUI have shown clearly that loops and plumes show ubiquitous transverse
motions \citep{anfinogentov2013, nakariakov2024}. Since the loop is displaced from its axis, these transverse motions are often identified as kink 
waves
\citep{vd2008,goossens2009}, in contrast to \alfven{} waves which are thought to have a torsional motion in a cylindrical structure. The transverse
waves manifest as the decayless
transverse waves \citep{tian2012,wang2012,nistico2013} in coronal loops. These loops are thought to be steadily fed with energy from its
convective footpoint motions \citep[e.g.][and references therein]{karampelas2021} in order to counteract the strong damping that is observed for
impulsively excited waves during flares \citep[e.g.][and references therein]{nechaeva2019}. They have a standing character
\citep{anfinogentov2015}, even though that is not so clear in very short loops \citep{gao2022,shrivastav2024}.\\
In coronal plumes or long coronal loops, the transverse waves are observed as propagating waves \citep{tomczyk2009,thurgood2014}. Also these
propagating waves are found to be damped \citep{tiwari2021}, although the weak damping rate seems to suggest only low density contrasts
\citep{morton2021}. It is unclear how these propagating kink waves are transformed higher up to \alfven{} waves \citep[as they should, as evidenced 
in PSP
observations, see e.g.][]{parashar2020} when the density contrast of the plumes is vanishing.\\
All these transverse waves must carry a significant amount of Poynting energy flux to the higher layers of the solar atmosphere. Energy 
estimates from observations go from a
few $\mathrm{W}/\mathrm{m}^2$ \citep{tomczyk2007,thurgood2014} to several $1000 \mathrm{W}/\mathrm{m}^2$ \citep{petrova2023}. While some of the
individual observed oscillations could counteract the heating requirements of quiet Sun or even active regions, the potential of wave heating with
transverse waves was only recently studied in a statistical way. \citet{lim2023} have performed a meta-analysis of transverse wave events
reported in the literature to show that their energy forms a power law of their frequency, with a supercritical slope. Thus, \citet{lim2023} found
that transverse waves can heat the corona when acting as an ensemble.

All these observational results motivated the development of 3D heating models for coronal loops and plumes \citep{vd2020ssrv}. These models rely on
the development of resonant absorption \citep[e.g.][]{goossens1992} kickstarting the Kelvin-Helmholtz instability \citep{terradas2008b,antolin2019}.
The KHI further develops under continued driving and encompasses the whole loop \citep{karampelas2018}. In some simulations, this mechanism was able
to balance the losses by optically thin radiative losses \citep{shi2021}, although that seems to be strongly dependent on the specific loop's
conditions \citep{demoortel2022}. Moreover, the heating provided by propagating transverse waves has been modeled in a section of the corona by
\citet{magyar2021}, and similar models for coronal heating by \alfven{} waves have been constructed by \citet{suzuki2005,shoda2019} and follow-up
works. Despite the promise that these numerical models hold, it has as yet not been possible to model the entire solar
atmosphere heated with these wave heating mechanisms, because of the required numerical resources. So, it is necessary and instructive to parametrise
the heating by transverse waves as 1D wave heating models and apply it to the full solar atmosphere. One example of this parametrisation is shown by
\citet{verdini2010}, who incorporate turbulent heating from \alfven{} waves and are able to explain the heating of plasma in a
expanding coronal hole and the acceleration of the solar wind.

More recently, other models driven by \alfven{} waves have been developed in 3D \citep[AWSOM by][]{vanderholst2014,downs2016,reville2020}. Such models
are being used all over the world for space weather forecasting. They are excellent at predicting the overall shape of the solar corona
\citep{riley2019}. Nowadays, similar models are being used to model stellar atmospheres
as well \citep[see e.g.][for some recent references]{alvarado2022,evensberget2024,cohen2024}. However, in these models, only the wave driving power
of the \alfven{} waves is taken into account. Moreover, the influence of background plasma structuring, except the structuring of the waves 
themselves (see $L_\perp$), perpendicular to the magnetic field on the wave dynamics has
been ignored, despite efforts from \citet{evans2012}. Here, we are motivated by the observations to also incorporate kink wave driving of the solar
and stellar atmospheres, following the approaches taken for the modelling of \alfven{} wave driving.

To accomplish this, we use the recent description for wave dynamics in terms of $Q$-variables \citep{vd2024}. These $Q$-variables allow to track
single waves in a co-propagating manner, and to include non-linear effects. The latter turns out to be important for propagating kink waves, because
they are known to show a turbulent evolution for a single wave pulse \citep{magyar2017}. This phenomenon has been named ``uniturbulence'', because
contrary to \alfven{} wave turbulence in a uniform background, only one unidirectional kink wave is needed to generate small scales. This has been
explained by
\citet{magyar2019,vd2020} by showing that kink waves naturally consist of both \elsasser{} variables simultaneously, which are constantly 
interacting. That
leads to damping of surface \alfven{} waves in numerical simulations and analytical theory \citep{ismayilli2022}. \\
\citet{vd2024} have provided the mathematical framework of the $Q$-variables and have rewritten the MHD equations in that formalism. We have
shown in that paper that the $Q$-variables allow a general wave perturbation to be split in its constituent wave modes. This would thus allow us to
complement the AWSOM idea (i.e. the MHD equations for the large scale evolution of the corona, complemented with \alfven{} wave evolution equations)
with an additional equation for the kink wave evolution. \\
In this paper, we first generalise the kink wave description to a flowing plasma as in the solar wind in Sec.~\ref{sec:kink}. Then we recapitulate 
the MHD equations in terms of $Q$-variables \citep[as found in][]{vd2024} in Sec.~\ref{sec:q}. From the combination of these 
sections, we derive a general energy evolution equation in the $Q$-formalism in Sec.~\ref{sec:ener_evol}. Then we
specialise this equation to the propagation and damping of kink waves (Sec.~\ref{sec:ener_ave}. Eventually, we formulate the set of equations for MHD 
plus \alfven{} and kink
evolution equations in Sec.~\ref{sec:mhd}, which we call the UAWSOM (uniturbulence and \alfven{} wave solar models) equations. Finally, we present 
the initial results of a
proof-of-concept implementation in a python package (Sec.~\ref{sec:numerics}).

\section{Kink wave variables and terminology}\label{sec:kink}
For the UAWSOM model, we consider the additional driving of kink waves in flux tubes, next to the \alfven{} waves. We consider the solar atmosphere 
filled with flux tubes, aligned with the magnetic field. The flux tubes we consider ought to model the coronal plumes which carry the main magnetic 
wave energy out into the solar wind. As such, in this section we first need to extend the model of \citet{vd2020} to also include the background 
flow. However, since many of the calculations are identical to the equations and steps in \citet{vd2020}, we only keep the bare minimum needed to get 
the required expressions for the calculations in Sec.~\ref{sec:energy}. 

To model kink waves in coronal plumes with background flow, we take an equilibrium configuration of a straight cylinder with homogeneous magnetic 
field $\vec{B}_0=B_0 \vec{e}_z$ and constant and uniform background flow of $\vec{V}=V\vec{e}_z$, without gravity, and no gas pressure $p=0$. In the 
first instance, we consider these equilibrium quantities not (or only weakly) to vary with height. We will relax this assumption in later sections, in 
a WKB-way. Our current configuration is most easily
described by the cylindrical coordinate system $(r,\varphi,z)$. We take a radial step function in density
\[\rho_0(r)=\begin{cases}
  	\rho_\mathrm{i} & \mbox{for } r\leq R,\\
  	\rho_\mathrm{e} & \mbox{for } r>R.
  \end{cases}
\]
The subscripts $i,e$ are associated with the interior and exterior region respectively. $R$ here is the radius of the coronal plume.
\par
As in \citet{vd2020}, we take the total pressure perturbation $P'=\frac{B_0\delta B_z}{\mu}$ as basic variable. We only consider propagating waves in
this
paper, which we describe with
\begin{equation}
	P'(r,\varphi,z,t)={\cal R}(r) \cos{\varphi} \cos{(k_z z-\omega t)}. \label{eq:p_prop}
\end{equation}
The function ${\cal R}(r)$ has the radial dependence of the pressure variation, for which we use the thin-tube limit $\delta=k_z R \ll 1$, and write
\begin{equation}
	{\cal R}(r)=\begin{cases} A\frac{r}{R} & \mbox{for } r\leq R\\ A\frac{R}{r} & \mbox{for } r>R\end{cases},
\end{equation}
with wave amplitude $A$. According to \citet{goossens1992}, in the thin-tube limit for a uniform flow, the wave frequency equals the kink frequency
\begin{equation}
    \omega=k_zV \pm \omega_\mathrm{k}, \quad \mbox{with} \quad \omega_\mathrm{k}=\sqrt{\frac{2k_z^2 B_0^2}{\mu (\rho_\mathrm{i}+\rho_\mathrm{e})}}=
\omega_\mathrm{Ae} \sqrt{\frac{2}{\zeta +1}},
\end{equation}
and where we have introduced the \alfven{} frequency $\omega_\mathrm{A}=k_zV_\mathrm{A}$ (in connection with the \alfven{} speed
$\vec{V}_\mathrm{A}=\vec{B}_0/\sqrt{\mu\rho}$, with magnetic permeability $\mu$) and the density contrast of the coronal plume as
$\zeta=\rho_\mathrm{i}/\rho_\mathrm{e}>1$, which we take to be greater than 1 always. \par

We compute the velocity components $(v_r,v_\varphi,v_z)$ using Eq.~10-11-12 of \citet{goossens1992}:
\begin{equation}
	\delta V_r=\frac{1}{\rho_0(\Omega^2-\omega_\mathrm{A}^2)} \frac{\partial }{\partial r}\frac{d P'}{dt}, \quad
\delta V_\varphi=\frac{1}{\rho_0(\Omega^2-\omega_\mathrm{A}^2)} \frac{1}{r}\frac{\partial }{\partial \varphi}\frac{d P'}{dt}, \quad \delta V_z=0,
\end{equation}
where the latter statement is true because of the cold plasma limit. In these equations, the total derivative $d/dt=\partial/\partial t +
V\partial/\partial z$ is the Lagrangian derivative. Moreover, $\Omega=\omega-k_zV$ is the Doppler-shifted frequency. The expression for the total
derivative is also the reason why we only consider propagating waves here \citep[in contrast to][]{vd2020}: for standing waves, the time derivative 
and $z$-derivative result in a
change of $z-t$ behaviour in both terms for standing waves.\\
With these notations, we can also relate the amplitude $A$ of the total pressure variations with the amplitude $\Upsilon$ of the velocity
perturbations generalising what we have obtained in \citet{vd2020}:
\begin{equation}
	\Upsilon = \frac{A}{R} \frac{\Omega}{\rho_0(\Omega^2-\omega_\mathrm{A}^2)}, \label{eq:amplitudes}
\end{equation}
where we now include the effect of background flow. 

In what follows, we will also need the expressions for the magnetic field perturbation $\vec{\delta B}$. For the remainder of this paper,
we will consider $\delta B_z=0$, because \citet{vd2020} have shown that its contributions are always higher order in $k_zR$ (even though we took
$P'=B_0\delta B_z/\mu\neq 0$ before!). For the other magnetic field components, we linearise the induction equation ($\partial \vec{B}_0/\partial t =
\nabla \times (\vec{V}\times \vec{B}_0)$) for a flowing plasma. The result for the components perpendicular to the magnetic field is
\begin{equation}
    \frac{\partial \vec{\delta B}_\perp}{\partial t} = B_0 \frac{\partial}{\partial z} \vec{\delta V}_\perp - V \frac{\partial}{\partial z}
\vec{\delta B}_\perp - \vec{\delta B}_\perp \nabla \cdot \vec{V}.
\end{equation}
In the last term, we left the dependence on the divergence of $\vec{V}$. In this cylindrical system, it is of course 0, but in the full model later
on, it will be non-zero. Still, we leave this term out here, because it is smaller (in a WKB sense) than the wave variations in the first two terms
on the RHS.

In the end, we generalise the expressions from \citet{vd2020} for the velocity and magnetic field perturbations, and we find that the 
generalisation is limited to replacing $\omega$ by the Doppler-shifted frequency $\Omega$:
\begin{align}
	\delta V_r &= \frac{\partial {\cal R}}{\partial r} \frac{\Omega}{\rho_0(\Omega^2-\omega_\mathrm{A}^2)} \cos{\varphi} \sin{(k_z z-\omega t)},
\label{eq:vr}\\
	\delta V_\varphi & = -\frac{{\cal R}}{r} \frac{\Omega}{\rho_0(\Omega^2-\omega_\mathrm{A}^2)} \sin{\varphi} \sin{(k_z z-\omega t)},
\label{eq:vphi}\\
	\delta B_r &= -\frac{\partial {\cal R}}{\partial r} \frac{k_zB_0}{\rho_0(\Omega^2-\omega_\mathrm{A}^2)} \cos{\varphi} \sin{(k_z z-\omega t)},\\
	\delta B_\varphi &= \frac{{\cal R}}{r} \frac{k_zB_0}{\rho_0(\Omega^2-\omega_\mathrm{A}^2)} \sin{\varphi}  \sin{(k_z z-\omega t)}, \\
	\delta B_z & = 0. \label{eq:bz}
\end{align}

In \citet{vd2020} \citep[and corresponding to the expression in][]{goossens2013}, we had computed with these variables (in a setting without flow)
that the wave energy density is given as
\begin{equation}
	\langle w \rangle = \pi R^2 \frac{\rho_\mathrm{i}+\rho_\mathrm{e}}{2}\Upsilon^2,
\end{equation}
which is averaged over the cross-section and the wavelength. Here, the wave energy $w$ is given as
\begin{equation}
	w^\pm = \frac{\rho_0 (\vec{\delta Z}^{\,\pm})^2}{4}, \label{eq:energy}
\end{equation}
and \begin{equation}\vec{\delta Z}^\pm=\vec{\delta V}\pm \frac{\vec{\delta B}}{\sqrt{\mu\rho_0}}\label{eq:elsasser}\end{equation}  is the
linearised version of the classical \elsasser{} variable $\vec{Z}=\vec{V}\pm \vec{V}_\mathrm{A}$. We then compute $w$ by
\begin{equation}
	w=w^++w^-.
\end{equation}
We sum both of these to cover the entirety of the wave energy, because the kink wave has both \elsasser{} components non-zero
\citep{magyar2019,vd2020}.

For the wave pressure $P_\mathrm{k}$, \citet{vd2020} found
\begin{align}
	P_\mathrm{ki}&= \frac{\omega_\mathrm{Ai}^2}{2\rho_\mathrm{i}(\Omega^2-\omega_\mathrm{Ai}^2)^2}\left(\frac{A}{R}\right)^2 \sin^2{(k_z z-\omega
t)}, \label{eq:pki}
\\
	P_\mathrm{ke}&= \frac{\omega_\mathrm{Ae}^2}{2\rho_\mathrm{e}(\Omega^2-\omega_\mathrm{Ae}^2)^2}\left[ \frac{A^2R^2}{r^4} \right] \sin^2{(k_z
z-\omega t)}, \label{eq:pke}
\end{align}
which we generalised immediately to have the Doppler-shifted frequency $\Omega$ instead.

From these generalisations of the kink wave functions of \citet{vd2020} to plumes that now include background flow, it seems that the influence of 
the flow is limited to replacing the wave frequency $\omega$ with the Doppler-shifted frequency $\Omega$ in all expressions. 

\section{$Q$-variables and terminology}\label{sec:q}
A second ingredient that is needed to formulate the UAWSOM model are the $Q$-variables. These were introduced in \citet{vd2024} as variables that are 
co-propagating variables with the waves's phase speeds. They are defined
as
\begin{equation}
	\vec{Q}^\pm=\vec{V}\pm \alpha \vec{B},
\end{equation}
where $\alpha$ is a parameter related to the phase speed of the wave. In \citet{vd2024} we have shown that these $Q$-variables are well suited to 
track any wave. Thus, we will use them to construct extra equations for propagating kink waves in plume regions, on top of the existing \alfven{} 
waves equations in the AWSOM model. Constructing these extra equations is possible, because $Q$-variables have as extra property that waves 
propagating in one
direction only have non-zero components in the co-propagating $Q$-component, generalising the convenience of the \elsasser{} variables for \alfven{}
waves to all other waves. We briefly list in this section the necessary notation and equations that form the basis for constructing the energy 
evolution equation for kink waves in coronal plumes. 

In our previous work \citet{vd2024}, we have shown that the MHD equations are rewritten in terms of $Q$-variables, with the
following form of the momentum and induction equation
\begin{multline}
			\frac{D^\mp }{Dt}\vec{Q}^\pm\mp\left(\frac{\vec{Q}^+-\vec{Q}^-}{4}\right)\frac{D^\mp }{Dt}\ln{\rho\alpha^2}=-v_\mathrm{s}^2 \nabla
\ln{\rho} - \frac{1}{8} \left(1-\frac{\Delta \alpha^2}{\alpha^2}\right)\nabla (\vec{Q}^+-\vec{Q}^-)^2 +
\frac{1}{4}\left(1-\frac{\Delta \alpha^2}{\alpha^2}\right)(\vec{Q}^+-\vec{Q}^-)^2\nabla \ln{\alpha} \\* -\frac{1}{4}\frac{\Delta \alpha^2}{\alpha^2}
(\vec{Q}^+-\vec{Q}^-)\cdot \nabla (\vec{Q}^+-\vec{Q}^-) + \frac{1}{4}\frac{\Delta \alpha^2}{\alpha^2} (\vec{Q}^+-\vec{Q}^-)
\nabla\cdot(\vec{Q}^+-\vec{Q}^-)
\\* \mp  \left(\frac{\vec{Q}^+-\vec{Q}^-}{8}\right) \nabla \cdot (3\vec{Q}^\pm-\vec{Q}^\mp) + \left(\frac{\vec{Q}^+-\vec{Q}^-}{4}\right)
\left(\left(\frac{\vec{Q}^+-\vec{Q}^-}{2}\right)\cdot\nabla \ln{\rho\alpha^2}\right). \label{eq:qmhd}
		\end{multline}
In this equation, we have employed the notation for the co-moving derivative $\frac{D^\pm}{Dt}=\frac{\partial}{\partial t}+\vec{Q}^\pm\cdot\nabla$,
the parameter $\Delta \alpha^2=\alpha^2-\frac{1}{\mu\rho}$ measures the deviation of the wave from an \alfven{} wave, and
$v_\mathrm{s}^2=\frac{\gamma
p}{\rho}$ is the square of the sound speed for a plasma with pressure $p$. Using the expression for the continuity relation
\begin{equation}
			\frac{D^\pm}{Dt} (\ln{\rho})= -\frac{1}{2}\nabla\cdot(3\vec{Q}^\pm-\vec{Q}^\mp)\pm
\frac{\vec{Q}^+-\vec{Q}^-}{2}\cdot\nabla\ln{\rho\alpha^2}, \label{eq:cont}
		\end{equation}
here we will use Eq.~\ref{eq:qmhd} in the form
\begin{multline}
    \frac{D^\mp }{Dt}\vec{Q}^\pm\mp\left(\frac{\vec{Q}^+-\vec{Q}^-}{4}\right)\frac{D^\mp }{Dt}\ln{\rho\alpha^2}=-v_\mathrm{s}^2 \nabla
\ln{\rho} - \frac{1}{8} \left(1-\frac{\Delta \alpha^2}{\alpha^2}\right)\nabla (\vec{Q}^+-\vec{Q}^-)^2 +
\frac{1}{4}\left(1-\frac{\Delta \alpha^2}{\alpha^2}\right)(\vec{Q}^+-\vec{Q}^-)^2\nabla \ln{\alpha} \\* -\frac{1}{4}\frac{\Delta \alpha^2}{\alpha^2}
(\vec{Q}^+-\vec{Q}^-)\cdot \nabla (\vec{Q}^+-\vec{Q}^-) + \frac{1}{4}\frac{\Delta \alpha^2}{\alpha^2} (\vec{Q}^+-\vec{Q}^-)
\nabla\cdot(\vec{Q}^+-\vec{Q}^-)
\\* \mp  \left(\frac{\vec{Q}^+-\vec{Q}^-}{8}\right) \left(-2 \frac{D^\pm}{Dt} (\ln{\rho}) \pm
(\vec{Q}^+-\vec{Q}^-)\cdot\nabla\ln{\rho\alpha^2}\right) + \left(\frac{\vec{Q}^+-\vec{Q}^-}{4}\right)
\left(\left(\frac{\vec{Q}^+-\vec{Q}^-}{2}\right)\cdot\nabla \ln{\rho\alpha^2}\right). \label{eq:qqmhd}
\end{multline}
For what follows, it is convenient to realise that
\begin{equation}
    \vec{Q}^+-\vec{Q}^-=2\alpha \vec{B} \label{eq:bfield}
\end{equation}
is in the direction of the magnetic field.

In our previous paper \citet{vd2024}, we have used these equations to describe surface \alfven{} waves. We have found that
\begin{equation}
	\alpha =\alpha_\mathrm{L}=\alpha_\mathrm{R}=\sqrt{\frac{2}{\mu\left(\rho_\mathrm{R}+\rho_\mathrm{L}\right)}}, \label{eq:alphakink}
\end{equation}
which also helps to recover the kink frequency $\omega_\mathrm{k}$ for such surface \alfven{} waves. Given the correspondence between surface
\alfven{} waves and kink waves \citep[especially in the long-wavelength limit,][]{goossens2012}, we will utilise these $\alpha$ values for kink waves
in cylinders (as in Sec.~\ref{sec:kink}) as well. We have previously shown that such a choice of $\alpha$ allows to decouple and filter out the waves 
of interest, in casu the kink waves. 

\section{Wave energy equations}\label{sec:wave_ener}
Now we can construct an energy evolution equation that will form the basis for the kink wave 
equation in the UAWSOM model. To that end, we start from the linearised $Q$-variable equations (Eq.~\ref{eq:qqmhd}) and transform it to an energy 
evolution equation. 

We consider how the wave energy evolves in a system which has the magnetic field $\vec{B}_0(z)$, the density 
$\rho_0(z)$ and the wind velocity $\vec{V}(z)$ as functions of $z$ and slowly evolving in time. 
The waves only have small perturbations on the background, and a classical
linearisation of the system is suitable. This is not true higher up in the solar wind (say above $\sim 3-5$ solar radii), but
\citep[following][]{vanderholst2014} we take this approximation nevertheless, because higher order (non-linear) terms are anyway included in the
system. Thus, we consider
\begin{equation}
    \vec{Q}^\pm=\vec{Q}_0^\pm+\vec{\delta Q}^\pm,\quad \mbox{and}\quad \ln{\rho} \approx \ln{\rho_0}+\delta R,\quad
 \ln{\rho
\alpha^2} \approx \delta R + \ln{\rho_0 \alpha^2},
\end{equation}
where we have introduced the notation $\delta R=\delta \rho/\rho_0$.
As in \citet{vd2024}, the phase speed variable $\alpha$ is not linearised in this process.
With the above linearisation, we obtain for the perpendicular components of $\vec{\delta Q}^\pm_\perp$:
\begin{multline}
    \frac{D^\mp}{Dt} \vec{\delta Q}^\pm_\perp + \vec{\delta Q}^\mp_\perp \cdot \nabla \vec{Q}^\pm_0 = -v_\mathrm{s}^2 \nabla \delta R \pm
\frac{\alpha}{2} \vec{\delta B}_\perp \frac{D^\mp}{Dt} \ln{\rho_0\alpha^2} \pm \frac{\alpha}{2} \vec{\delta B}_\perp \frac{D^\mp}{Dt} \delta R
-\Delta \alpha^2 \vec{B}_0 \cdot \nabla \vec{\delta B}_\perp - \Delta \alpha^2 \vec{\delta B}_\perp \cdot \nabla \vec{\delta B}_\perp\\ -
\frac{1}{\rho_0}\nabla P_\mathrm{w} \mp \left(\frac{\vec{\delta
Q}^+_\perp - \vec{\delta Q}^-_\perp}{8}\right)\left[ -2 \frac{D^\mp}{Dt} \ln{\rho_0} - 2 \frac{D^\mp}{Dt} \delta R \pm (\vec{\delta Q}^+_\perp -
\vec{\delta Q}^-_\perp)\cdot\nabla \ln{\rho_0\alpha^2} \mp (\vec{\delta Q}^+_\perp - \vec{\delta Q}^-_\perp)\cdot \nabla \delta R\right. \\ \left.
\pm
(\vec{Q}^+_0-\vec{Q}^-_0)\cdot \nabla \ln{\rho_0\alpha^2}\right] \\ +\left(\frac{\vec{\delta Q}^+_\perp - \vec{\delta Q}^-_\perp}{8}\right)
\left[(\vec{Q}^+_0-\vec{Q}^-_0)\cdot \nabla \ln{\rho_0\alpha^2} + (\vec{\delta Q}^+_\perp - \vec{\delta Q}^-_\perp)\cdot\nabla \ln{\rho_0\alpha^2} +
(\vec{\delta Q}^+_\perp - \vec{\delta Q}^-_\perp)\cdot \nabla \delta R\right],
\end{multline}
where we have conveniently left some terms in $\vec{\delta B}_\perp$ to shorten the notation. $P_\mathrm{w}=\frac{\delta B^2}{2\mu}$ is the wave
pressure in the equation. It
will become either the kink wave pressure $P_\mathrm{k}$ or the \alfven{} wave pressure $P_\mathrm{A}$.

Because the kink wave is nearly incompressible, we take all terms with $\delta R$ to be zero, since the kink wave's compression scales with
$k_z^2R^2$. Moreover, we assume that the perpendicular variations of equilibrium quantities (e.g. $\vec{\delta Q}^\mp_\perp \cdot \nabla
\vec{Q}^\pm_0$) are also 0. This is in principle incorrect, because in the kink wave model of Sec.~\ref{sec:kink} large variations of density are
present across the magnetic field. However, we will account for the effect of those gradients only through the use of $\alpha$ and averaging across
the cross-section of the coronal plume. The resulting equation is then
\begin{multline}
    \frac{D^\mp}{Dt} \vec{\delta Q}^\pm_\perp =  \pm
\frac{\alpha}{2} \vec{\delta B}_\perp \frac{D^\mp}{Dt} \ln{\rho_0\alpha^2}
-\Delta \alpha^2 \vec{B}_0 \cdot \nabla \vec{\delta B}_\perp - \Delta \alpha^2 \vec{\delta B}_\perp \cdot \nabla \vec{\delta B}_\perp\\ -
\frac{1}{\rho_0}\nabla P_\mathrm{w} \mp \left(\frac{\vec{\delta
Q}^+_\perp - \vec{\delta Q}^-_\perp}{8}\right)\left[ -2 \frac{D^\mp}{Dt} \ln{\rho_0}
\pm
(\vec{Q}^+_0-\vec{Q}^-_0)\cdot \nabla \ln{\rho_0\alpha^2}\right] +\left(\frac{\vec{\delta Q}^+_\perp - \vec{\delta Q}^-_\perp}{8}\right)
\left[(\vec{Q}^+_0-\vec{Q}^-_0)\cdot \nabla \ln{\rho_0\alpha^2}\right]. \label{eq:deltaq}
\end{multline}
It is sometimes convenient to take the limit of $\alpha^2\to \frac{1}{\mu\rho}, \Delta \alpha^2\to 0, \vec{\delta Q}^\pm_\perp \to \vec{\delta Z}^\pm$
to
check the results. Here we can use this limit to see that the previous equation coincides in that limit with Eq.~20 of \citet{vanderholst2014}. Since 
we have
neglected perpendicular gradients of equilibrium quantities, we do not have their terms with gradients of $\vec{u}$ and $\vec{B}$, but we did keep
the term with the wave pressure.

The linear portion of the left term in Eq.~\ref{eq:deltaq} is
\begin{equation}
    \frac{\partial}{\partial t} \vec{\delta Q}^\pm_\perp + \vec{Q}^\mp_0\cdot \nabla \vec{\delta Q}^\pm_\perp,
\end{equation}
which shows that the wave propagates the speed $\vec{Q}^\mp_0=\vec{V}\mp \alpha\vec{B}_0$. This indicates at the same time that $\alpha
\vec{B}_0$ is the phase speed of the wave, and also that $\vec{\delta Q}^+_\perp$ belongs to the downward propagating wave (if the magnetic field is
pointing up).

\subsection{Energy evolution}\label{sec:ener_evol}
Following the procedure of \citet{vanderholst2014}, we now multiply Eq.~\ref{eq:deltaq} with $\rho_0 \vec{\delta Q}^\pm_\perp/2$. Then we have an 
evolution equation for the $Q$-density $W^\pm$:
\begin{equation}
    W^\pm = \frac{\rho_0}{4} \left(\vec{\delta Q}^\pm_\perp\right)^2. \label{eq:qdensity}
\end{equation}
This expression is seemingly different than the wave energy density $w$ (Eq.~\ref{eq:energy}), but it is only an apparent difference. In
appendix~\ref{sec:energy}, we show that
\begin{equation}
    W^\pm = \rho_0 \delta V^2_\perp,
\end{equation}
and thus that the $Q$-density is equal to the wave energy density $w$ if the wave energy is in equipartition.

We will now use this $Q$-density to construct an energy evolution equation for the kink waves. As stated before, we multiply Eq.~\ref{eq:deltaq}
with $\rho_0 \vec{\delta Q}^\pm_\perp/2$. We then obtain
\begin{multline}
    \frac{D^\mp}{Dt} W^\pm - W^\pm
\frac{D^\mp}{Dt}\ln{\rho_0} -\frac{1}{2} \left(W^\pm - CC\right) \frac{D^\pm}{Dt}\ln{\rho_0}= -\Delta \alpha^2 \frac{\rho_0}{2} \vec{\delta
Q}^\pm_\perp
\cdot \left(\vec{B}_0 \cdot \nabla \vec{\delta B}_\perp\right) - \Delta \alpha^2 \frac{\rho_0}{2}\vec{\delta Q}^\pm_\perp \left(\vec{\delta B}_\perp
\cdot
\nabla \vec{\delta B}_\perp\right) +\frac{1}{2} \left(W^\pm-CC\right) \frac{D^\mp}{Dt}\ln{\rho_0\alpha^2},
\end{multline}
where we have introduced the shorthand $CC=\rho_0\frac{\vec{\delta Q}^+_\perp\cdot \vec{\delta Q}^-_\perp}{4}$ for the cross-correlation between
upward and downward propagating waves. We grouped the terms not involving $\Delta \alpha^2$ or $\ln{\rho_0\alpha^2}$ (basically the effects of
\alfven{} waves, as was already incorporated in AWSOM) to the left-hand side. The term with the wave pressure was dropped at this stage, because
(1) its net contribution is in the $z$-direction, and (2) the time and spatially averaged term is added to the momentum equation of the background
(Eq.~\ref{eq:mom}). In any case, the radial contribution of the pressure term has been averaged out previously. \\ We rewrite the left-hand side of
the previous equation following the AWSOM lead of
\citet{vanderholst2014}, using the equation for the conservation of mass of the background plasma
\begin{equation}
    \frac{\partial}{\partial t}\ln{\rho_0} + \vec{V}\cdot\nabla \ln{\rho_0} + \nabla\cdot \vec{V}=0,
\end{equation}
to obtain
\begin{multline}
    \frac{D^\mp}{Dt} W^\pm - W^\pm
\frac{D^\mp}{Dt}\ln{\rho_0} -\frac{1}{2} \left(W^\pm - CC\right) \frac{D^\pm}{Dt}\ln{\rho_0} =\\ \frac{\partial W^\pm}{\partial t} + \nabla
\cdot((\vec{Q}^\mp_0+\vec{\delta Q}^\mp_\perp)W^\pm) + \frac{1}{2} CC \frac{D^\pm \ln{\rho_0}}{Dt} + W^\pm \left( \frac{1}{2} \nabla\cdot \vec{V} \pm
\frac{1}{2} \alpha\vec{B}_0\cdot \nabla\ln{\rho_0\alpha^2}\right),
\end{multline}
where we have used once again the incompressibility of the wave mode and perpendicular gradients of background quantities to be zero. Finally, we
obtain as evolution equation for the $Q$-density
\begin{multline}
    \frac{\partial W^\pm}{\partial t} + \nabla
\cdot((\vec{Q}^\mp_0+\vec{\delta Q}^\mp_\perp)W^\pm) + \frac{1}{2} CC \frac{D^\pm \ln{\rho_0}}{Dt} +  \frac{W^\pm}{2} \nabla\cdot \vec{V} =
\frac{W^\pm}{2} \left(\frac{D^\mp}{Dt} \mp \alpha\vec{B}_0\cdot \nabla\right) \ln{\rho_0\alpha^2} - \frac{CC}{2}
\frac{D^\mp}{Dt}\ln{\rho_0\alpha^2}\\ -\Delta \alpha^2 \frac{\rho_0}{2} \vec{\delta Q}^\pm_\perp
\cdot \left(\vec{B}_0 \cdot \nabla \vec{\delta B}_\perp\right) - \Delta \alpha^2 \frac{\rho_0}{2}\vec{\delta Q}^\pm_\perp\cdot \left(\vec{\delta
B}_\perp
\cdot
\nabla \vec{\delta B}_\perp\right). \label{eq:wqevol}
\end{multline}
The second term in the right-hand side of this equation in principle groups with the third term on the left-hand side. However, this way of writing 
is more elegant, because these right-hand side terms describe how the wave energy density evolution is modified if $\alpha\neq 1/\sqrt{\mu\rho}$, 
i.e. a different wave is considered than an \alfven{} wave. \\
The rightmost term in this equation describes the energy dissipation by uniturbulence. There is self-interaction of the wave (any wave!) if
$\Delta \alpha^2\neq 0$, leading to a cascade of energy in a turbulent way \citep{ismayilli2024}.

\subsubsection{Characterisation of the uniturbulence term}\label{sec:uniterm}

Let us further specialise these equations for single kink waves, i.e. only taking into account the effect of the kink waves on themselves. The terms
with $\ln{\rho_0\alpha^2}$ will cancel out, because the $\alpha$ for the kink wave (Eq.~\ref{eq:alphakink}) shows that it is still proportional to
the density (interior or external does not matter here). The only variation in that term would be due to spatial and temporal variations of the
density contrast $\zeta$, but we will take these to be 0 in this paper, assuming that the density contrast and filling factor remain constant in the
entire domain. Later on, in Sec.~\ref{sec:numerics}, the parameter $\zeta$ will in fact change with height, but its change is very small indeed, so 
the 
approximation here is fine. 

For working out the uniturbulence terms, we consider that there is only one of $\vec{\delta Q}^\pm_\perp$, in order to have only the
self-interaction \citep{magyar2019}. Then we can use Eq.~\ref{eq:bfield} to write that
\begin{equation}
    \vec{\delta Q}^\pm_\perp = \pm 2 \alpha \vec{\delta B}_\perp. \label{eq:dqbperp}
\end{equation}
With this equation, we can rewrite
\begin{equation}
    - \Delta \alpha^2 \frac{\rho_0}{2}\vec{\delta Q}^\pm_\perp \left(\vec{\delta B}_\perp \cdot
\nabla \vec{\delta B}_\perp\right) = \mp  \Delta \alpha^2 \mu \rho_0 \alpha \vec{\delta B}_\perp \cdot \nabla P_\mathrm{k}.
\end{equation}
In the exterior region of the plume, the latter expression reduces to
\begin{equation}
    \mp \frac{1-\zeta}{1+\zeta} \alpha \vec{\delta B}_\perp \cdot
\nabla P_\mathrm{k}, \label{eq:extpk}
\end{equation}
where we have used the expression for $\alpha$ from Eq.~\ref{eq:alphakink}. Using Eq.~\ref{eq:vr}-\ref{eq:bz} \citep[or Eq.~40 \& 44 in][]{vd2020},
we have that $\vec{\delta B}_{\perp\mathrm{i}} \cdot
\nabla P_\mathrm{ki}=0$ up to leading orders of $k_zR$. The remaining contribution for uniturbulent damping is
\begin{equation}
    \vec{\delta B}_{\perp \mathrm{e}}\cdot \nabla P_\mathrm{ke} = -2\frac{A^3R^3}{r^7} \frac{k_z B_0
\omega_\mathrm{Ae}^2}{\rho_\mathrm{e}^2(\Omega^2-\omega_\mathrm{Ae}^2)^3}\cos{\varphi} \sin^3{(k_zz-\omega t)}. \label{eq:bgradp}
\end{equation}
For the radial averaging, we integrate from $R$ to $\gamma R$ as for the energy \citep[as we have also done in][see also the procedure in
Sec.~\ref{sec:energy} for an example]{goossens2013,vd2014,vd2020}. For that integral we have
\begin{equation}
    \int_R^{\gamma R} \frac{1}{r^7} rdr = \frac{1}{5R^5}\left(\frac{\gamma^5-1}{\gamma^5}\right) = \frac{1}{5R^5} (1-f^{5/2}).
\end{equation}
As was discussed in \citet{vd2020}, the expression Eq.~\ref{eq:bgradp} is an odd-powered periodic function in $\varphi$ and time, and its average
would result in 0. Thus, it makes sense to take the RMS average of this term to quantify the uniturbulent damping. The physical interpretation is
that all small scales are cascaded to even smaller scales before they experience the inverse cascade in the second half of the period. From
\citet{vd2020}, we know that the RMS average will yield a $\sqrt{\pi}$ from the integral over $\varphi$, and the RMS average over $t$ yields a factor
$\sqrt{5/16}$. We then obtain as result
\begin{equation}
    -\frac{\zeta-1}{\zeta+1} \sqrt{\pi} \sqrt{\frac{5}{16}} \frac{2A^3}{5R^2} \frac{\alpha k_z B_0
\omega_\mathrm{Ae}^2}{\rho_\mathrm{e}^2(\Omega^2-\omega_\mathrm{Ae}^2)^3}(1-f^{5/2}),
\end{equation}
where we have kept the minus sign only to extract energy from the equation. Further substituting the relevant expressions for $\Upsilon$
(Eq.~\ref{eq:amplitudes}), we then write
\begin{equation}
    -\frac{\zeta-1}{\zeta+1} \sqrt{\frac{\pi}{20}} \alpha k_z B_0 R \frac{\rho_\mathrm{e} \omega_\mathrm{Ae}^2 \Upsilon^3}{\Omega^3} (1-f^{5/2}),
\end{equation}
which is further simplified with the expression for $\Omega=\alpha k_z B_0$ to
\begin{equation}
    -\frac{\zeta-1}{\zeta+1} \sqrt{\frac{\pi}{20}} R \frac{\rho_\mathrm{e} \omega_\mathrm{Ae}^2 \Upsilon^3}{\Omega^2} (1-f^{5/2}).
\end{equation}
Here we use once again that $\Omega=\pm\alpha k_zB_0=\pm \omega_\mathrm{Ae}\sqrt{\frac{2}{\zeta+1}}$ to further simplify to
\begin{equation}
    - \frac{\zeta-1}{4} \sqrt{\frac{\pi}{5}} R \rho_\mathrm{e} \Upsilon^3 (1-f^{5/2}).
\end{equation}
This expression is equal to the damping rate found in \citet{vd2020}, confirming the earlier result with the use of the new $Q$-variables.
For the implementation in a numerical code, it is most convenient to write this term as an expression of the cross-section and wavelength-averaged
energy density $\langle W^\pm \rangle$ \citep[see][and Sec.~\ref{sec:energy} for more details]{vd2014,vd2020}:
\begin{equation}
    \langle W^\pm \rangle = \frac{k_z}{2\pi} \int_z dz \int_\varphi d\varphi \int_0^{\gamma R} W^\pm r dr = \pi R^2 \Upsilon^2
\frac{\rho_\mathrm{i}+\rho_\mathrm{e}(1-f)}{2}, \label{eq:cross-average}
\end{equation}
to
\begin{equation}
    - \langle W^\pm \rangle^{3/2} \frac{1}{\pi R^2} \frac{1}{\sqrt{\rho_\mathrm{e}}}  \sqrt{\frac{2}{5}}
\frac{\zeta - 1}{2} \frac{1-f^{5/2}}{(\zeta+1-f)^{3/2}}. \label{eq:unidamping}
\end{equation}

\subsubsection{Characterisation of the $\vec{B}_0\cdot\nabla \vec{\delta B}_\perp$ term}
In the previous subsection, we have simplified the energy cascade term for the uniturbulence damping mechanism. We will now simplify the remaining, 
unknown terms in Eq.~\ref{eq:wqevol}. To that end, we use Eq.~\ref{eq:dqbperp} to simplify the term in Eq.~\ref{eq:wqevol} containing 
$\vec{B}_0\cdot\nabla \vec{\delta B}_\perp$:
\begin{equation}
    -\Delta \alpha^2 \frac{\rho_0}{2} \vec{\delta Q}^\pm_\perp
\cdot \left(\vec{B}_0 \cdot \nabla \vec{\delta B}_\perp\right)= \mp \Delta \alpha^2 \frac{\rho_0}{2} 2\alpha \vec{\delta B}_\perp \cdot
\left(\vec{B}_0 \cdot
\nabla \vec{\delta B}_\perp\right) = \mp \Delta \alpha^2 \mu\rho_0 \alpha \vec{B}_0 \cdot \nabla \left(\frac{\vec{\delta B}_\perp^2}{2\mu}\right) =
\mp
\Delta \alpha^2 \mu\rho_0 \alpha \vec{B}_0 \cdot \nabla P_\mathrm{k}.
\end{equation}
We have previously (Eq.~\ref{eq:extpk}) used that in the region exterior to the plume we have as value of $\mu\rho_e\Delta \alpha_\mathrm{e}^2 =
\frac{1-\zeta}{1+\zeta}$. However, to quantify the term in question, we also need the expression in the interior region:
\begin{equation} \mu\rho_\mathrm{i}\Delta \alpha_\mathrm{i}^2 = \frac{\zeta - 1}{\zeta+1}, \label{eq:intpk} \end{equation}
i.e. it is the opposite of the value in the exterior.\par
Then we need to average the contributing term over the entire cross-section, as we have done for the uniturbulence term in Subsec.~\ref{sec:uniterm}.
We need evaluate
\begin{equation}
    \mp \left\lbrace \int_0^R rdr \mu\rho_\mathrm{i}\Delta \alpha_\mathrm{i}^2 \alpha\vec{B}_0\cdot\nabla P_\mathrm{ki} + \int_R^{\gamma R} rdr
\mu\rho_\mathrm{e}\Delta \alpha_\mathrm{e}^2 \alpha\vec{B}_0\cdot\nabla P_\mathrm{ke}\right\rbrace.
\end{equation}
Using the results of Eq.~\ref{eq:extpk} and \ref{eq:intpk}, we then have
\begin{equation}
    \mp \left( \frac{\zeta-1}{\zeta+1}\right) \left\lbrace \int_0^R rdr \alpha\vec{B}_0\cdot\nabla
P_\mathrm{ki} - \int_R^{\gamma R} rdr
\alpha\vec{B}_0\cdot\nabla P_\mathrm{ke}\right\rbrace.
\end{equation}
Now we use the fact that the field-aligned gradient will be equal for the internal and external pressure in the WKB approximation, so it can be moved
out of the integrals. We then are left to calculate
\begin{equation}
    \mp \left( \frac{\zeta-1}{\zeta+1}\right) \alpha\vec{B}_0\cdot\nabla \left\lbrace \int_0^R P_\mathrm{ki}rdr
 - \int_R^{\gamma R}  P_\mathrm{ke} rdr\right\rbrace.
\end{equation}
With the aid of the expressions~\ref{eq:pki} and \ref{eq:pke}, and utilising Eqs. 43 and 46 in \citet{vd2020}, we obtain
\begin{equation}
    \mp \left( \frac{\zeta-1}{\zeta+1}\right) \alpha\vec{B}_0\cdot\nabla \left\lbrace \frac{\Upsilon^2 \pi
R^2}{2\omega^2}(\rho_\mathrm{i}\omega_\mathrm{Ai}^2-\rho_\mathrm{e}\omega_\mathrm{Ae}^2)(1-f)\right\rbrace. \label{eq:BgradP}
\end{equation}
When we take the filling factor to be small (i.e. $f\ll 1$), then the term in the curly brackets is exactly 0.

\subsubsection{Evolution equation for cross-sectionally averaged wave energy}\label{sec:ener_ave}
Now we have simplified all terms in the energy evolution equation (Eq.~\ref{eq:wqevol}). We can now gather all the previous terms and their 
simplified expression to achieve our goal of obtaining an evolution equation for the kink wave energy density. 
With the expressions \ref{eq:unidamping} and \ref{eq:BgradP}, we may write the final equation for the kink wave energy evolution as
\begin{equation}
        \frac{\partial \langle W^\pm \rangle}{\partial t} + \nabla
\cdot(\vec{Q}^\mp_0\langle W^\pm \rangle) +  \frac{\langle W^\pm \rangle}{2} \nabla\cdot \vec{V} =
 - \langle W^\pm \rangle^{3/2} \frac{1}{\pi R^2} \frac{1}{\sqrt{\rho_\mathrm{e}}}
\sqrt{\frac{2}{5}}
\frac{\zeta - 1}{2} \frac{1-f^{5/2}}{(\zeta+1-f)^{3/2}}.
\end{equation}
This equation has inconvenient units to add up with the traditional expression for the \alfven{} wave energy density, since $\langle W^\pm\rangle$
has been integrated over the cross-section of the magnetic flux tube. Thus, it is more convenient to once again normalise it to the relevant
cross-sectional area $\pi \gamma^2R^2$. Thus, we follow the procedure of \citet{goossens2013} to compute the kink wave energy density. We define
\begin{equation}
    W_\mathrm{k}^\pm=\frac{\langle W^\pm\rangle}{\pi \gamma^2R^2}
\end{equation}
as the cross-sectionally averaged energy density of the kink waves, to obtain as evolution equation for the kink wave energy density
\begin{equation}
    \frac{\partial W_\mathrm{k}^\pm}{\partial t} + \nabla
\cdot(\vec{Q}^\mp_0W_\mathrm{k}^\pm) +  \frac{W_\mathrm{k}^\pm}{2} \nabla\cdot \vec{V} =
 - \frac{1}{L_{\perp,\mathrm{VD}}} \frac{1}{\sqrt{\rho_\mathrm{e}}}
(W_\mathrm{k}^\pm)^{3/2} , \label{eq:unidiss}
\end{equation}
where we have defined the appropriate length for the damping of the waves as
\begin{equation}
    L_{\perp,\mathrm{VD}}(R,\zeta,f) = \left(\frac{1}{R\sqrt{f\pi}}
\sqrt{\frac{2}{5}}
\frac{\zeta - 1}{2} \frac{1-f^{5/2}}{(\zeta+1-f)^{3/2}}\right)^{-1}.
\end{equation}
One may say that the introduction of such a length is not any better than the infamous $L_{\perp}$ parameter for the turbulent damping of \alfven{}
waves. However, the difference is that the \alfven{} turbulence parameter is solely based on a phenomenological approach, but here we offer a full 
description of it for uniturbulence heating.
Given that $R$, $f$, and $\zeta$ all vary with distance along a magnetic field line, it is clear that also $L_{\perp,\mathrm{VD}}$ will vary with
height in the atmosphere. A reasonable prescription for this parameter may be obtained from initial atmospheric models \citep[e.g.][]{sishtla2022}. 
In the atmospheric model we use in Sec.~\ref{sec:numerics}, the value ranges from 6.3Mm-7.3Mm, which is similar to values for \alfven{} waves 
\citep[e.g.][and references therein]{sharma2023}.

As an alternative, one may consider the theory of \citet{hillier2019} to quantify the energy cascade rate in the saturated regime. In that case, a
new expression would emerge, but it would come down to replacing $L_{\perp,\mathrm{Hillier}}$ in the previous equation.

To Eq.~\ref{eq:wqevol} an ad-hoc reflection term should be added. We name these terms ${\cal R}_\mathrm{A,k}$ for the respective waves. One could
take the same reflection as in \citet{vanderholst2014} for the \alfven{}
and kink waves alike. Or we may consider the reflection expression from \citet{reville2020}. One could also opt for taking a different reflection of
the \alfven{} and kink waves.

\section{Energy equations for the background}\label{sec:mhd}
Now we need to incorporate the kink wave energy evolution equation (Eq.~\ref{eq:unidiss}) into the larger system of AWSOM equations. For this, we 
need to model the slow background variation through regular MHD, but we supplement the momentum equation with the forces exerted by the wave 
pressures
$P_\mathrm{k}$ and $P_\mathrm{A}$. To express the conservation of energy, we need to compute the equation for the conservation of wave energy. To
that end, we sum Eq.~\ref{eq:wqevol} for the upward and downward \alfven{} wave, and Eq.~\ref{eq:unidiss} for the upward and downward kink wave. Once
again following
\citet{vanderholst2014}, we also take $CC=0$. We obtain
\begin{multline}
    \frac{\partial }{\partial
t}(W^+_\mathrm{A}+W^-_\mathrm{A}+W^+_\mathrm{k}+W^-_\mathrm{k})+\nabla\cdot(\vec{Z}_0^-W^+_\mathrm{A}+\vec{Z}_0^+W^-_\mathrm{A}+\vec{Q}_0^-W^+_\mathrm
{k}+\vec{Q}_0^+W^-_\mathrm{k})+\frac{1}{2}(W^+_\mathrm{A}+W^-_\mathrm{A}+W^+_\mathrm{k}+W^-_\mathrm{k}) \nabla\cdot \vec{V}_0=\\
 - \Gamma^+W^+_\mathrm{A}-\Gamma^-W^-_\mathrm{A}- \frac{1}{L_{\perp,\mathrm{VD}}} \frac{1}{\sqrt{\rho_\mathrm{e}}} (
(W^+_\mathrm{k})^{3/2}+ (W^-_\mathrm{k})^{3/2}), \label{eq:waveenergy}
\end{multline}
where $\vec{Z}_0^\pm=\vec{V}_0\pm \vec{V}_\mathrm{A}$ for the \alfven{} waves and $\vec{Q}_0^\pm$ is the kink phase speed in the moving frame. The
term containing the uniturbulent damping is redefined to have the cross-sectionally averaged wave energy density. The terms with $\Gamma^\pm$ are the
terms of the \alfven{} wave turbulence \citep{vanderholst2014} containing the infamous $L_\perp$ parameter. These terms originate in the $\nabla
\cdot \delta \vec{Q}^\mp_\perp W^\pm$ terms of Eq.~\ref{eq:wqevol} that express the cascade due to interaction with counterpropagating waves. For
now, we have assumed that
counterpropagating kink waves do not interact (and thus that the corresponding $\Gamma$ terms are 0). Indeed, it is likely that the cascade due to
their interaction is much smaller than the uniturbulent
self-cascade (Eq.~\ref{eq:unidiss}). The last term will be combined with the terms for the work of the wave pressure
forces, using the
expression in \citet{vanderholst2014}:
\begin{equation}
    P_\mathrm{A}=\frac{W^+_\mathrm{A}+W^-_\mathrm{A}}{2}, \quad \mbox{and} \quad
P_\mathrm{k}=\frac{W^+_\mathrm{k}+W^-_\mathrm{k}}{2\mu\rho_0\alpha^2}.
\end{equation}
The terms in the right-hand side of Eq.~\ref{eq:waveenergy} should not be taken into account into the integrated energy equation of MHD plus the
waves. Indeed, the cascade energy (beit via \alfven{} wave turbulence or uniturbulence) is added into the system as heat. Thus, those damping terms
would cancel out with the heating terms.

We have also taken all cross-interactions to be zero. For instance terms like $\vec{\delta Z}^+\cdot \nabla W^+_\mathrm{k}$. We could
make some Fourier based arguments why these will not contribute, but \citet{guo2019b} have already shown that the presence of \alfven{} waves does in
fact facilitate the cascade of (admittedly standing) kink waves.

Let us now formulate the conservation equations for the entire system. We start with the conservation of mass and the induction equation:
\begin{align}
    \frac{\partial \rho_0}{\partial t} + \nabla\cdot (\rho_0\vec{V})=0, & \\
    \frac{\partial \vec{B}_0}{\partial t} - \nabla\times(\vec{V}\times\vec{B}_0) = 0, & \label{eq:induction}
\end{align}
which are unchanged from straight MHD. Then, we formulate the conservation of momentum
\begin{equation}
    \frac{\partial \rho_0 \vec{V}}{\partial t} + \nabla\cdot(\rho_0\vec{V}\vec{V}-\frac{1}{\mu}\vec{B}_0\vec{B}_0) + \nabla(p+\frac{B_0^2}{2\mu} +
P_\mathrm{A}+P_\mathrm{k}) = - \rho_0\vec{g}(\vec{r}),\label{eq:mom}
\end{equation}
where additionally to the traditional terms (including the gravity $\vec{g}$, depending on the radial distance from the Sun), we have included the
wave pressures $P_\mathrm{A}$ and $P_\mathrm{k}$.\\
As the equation for internal energy, we take
\begin{equation}
    \frac{\partial}{\partial t}\frac{p}{\gamma -1} + \nabla \cdot \frac{p\vec{V}}{\gamma -1} + p\nabla\cdot \vec{V} = -{\cal
L}+{\cal H}_{\mathrm{A}+\mathrm{k}},
\label{eq:gasp}
\end{equation}
where we have incorporated radiative losses ${\cal L}$ and the heating ${\cal H}_{\mathrm{A}+\mathrm{k}}=
\Gamma^+W^+_\mathrm{A}+\Gamma^-W^-_\mathrm{A}+  \frac{1}{L_{\perp,\mathrm{VD}}(R,\zeta,f)} \frac{1}{\sqrt{\rho_\mathrm{e}}} (
(W^+_\mathrm{k})^{3/2}+ (W^-_\mathrm{k})^{3/2})$ by the \alfven{} and kink waves.
To obtain the final energy equation, we add Eq.~\ref{eq:waveenergy}, Eq.~\ref{eq:gasp},
$\vec{V}\cdot$Eq.~\ref{eq:mom} and $\vec{B}\cdot$Eq.~\ref{eq:induction}, resulting in
\begin{multline}
    \frac{\partial}{\partial t}\left(\rho_0 \frac{V^2}{2}+\frac{p}{\gamma
-1}+\frac{B_0^2}{2\mu}+\sum W_\mathrm{A,k}^\pm\right)+\nabla\cdot\left(\left[\rho_0\frac{V^2}{2}+\frac{p}{\gamma
-1}+\frac{B_0^2}{2\mu}\right]\vec{V}-\vec{B}_0\frac{\vec{V}\cdot\vec{B_0}}{\mu}\right)\\ +
\nabla\cdot\left(\vec{Z}_0^-W^+_\mathrm{A}+\vec{Z}_0^+W^-_\mathrm{A}+\vec{Q}_0^-W^+_\mathrm
{k}+\vec{Q}_0^+W^-_\mathrm{k}\right)+\nabla\cdot\left(\left[P_\mathrm{A}+
P_\mathrm{k}\right]\vec{V}\right)+(\mu\rho_0\alpha^2-1)P_\mathrm{k}\nabla\cdot\vec{V}=-{\cal
L} - \rho_0\vec{V}\cdot\vec{g}(\vec{r})
\end{multline}
where we have used the shorthand $\sum W_\mathrm{A,k}^\pm=W_\mathrm{A}^++W_\mathrm{A}^-+W_\mathrm{k}^++W_\mathrm{k}^-$. Moreover, we have used the
expression for $P_\mathrm{k}$ in terms of the $Q$-densities:
\begin{equation}
    P_\mathrm{k}=\frac{\delta
B^2}{2\mu}=\frac{1}{\mu\rho_0\alpha^2}\frac{W^+_\mathrm{k}+W^-_\mathrm{k}}{2}=\frac{1+\zeta}{4}\left(W^+_\mathrm{k}+W^-_\mathrm{k}\right),
\end{equation}
in the case that $CC=0$. The factor $(\mu\rho_0\alpha^2-1)$ is present in this equation, because of the non-coincidence between the wave pressure 
$P_\mathrm{k}$ and the average of the wave energies. It reduces to $\frac{1-\zeta}{1+\zeta}$ for kink waves. For an \alfven{} wave, this term reduces 
to 0, because the factor  $(\mu\rho_0\alpha^2-1)$ is zero when $\alpha^2=1/\mu\rho_0$, i.e. for an Alfv\'en wave. The resulting extra term contains 
the density
contrast, which is varying with height in the solar
atmosphere. Thus, also here, a model parameter will appear in the equations which needs to be fixed ad-hoc, just like $L_{\perp,\mathrm{VD}}$.

Finally, the full set of equations can be written as
\begin{align}
     \frac{\partial \rho_0}{\partial t} + \nabla\cdot (\rho_0\vec{V}) &=0,  \label{eq:continuity}\\
    \frac{\partial \vec{B}_0}{\partial t} - \nabla\times(\vec{V}\times\vec{B}_0) & = 0,  \\
    \frac{\partial \rho_0 \vec{V}}{\partial t} + \nabla\cdot(\rho_0\vec{V}\vec{V}-\frac{1}{\mu}\vec{B}_0\vec{B}_0) + \nabla(p+\frac{B_0^2}{2\mu} +
P_\mathrm{A}+P_\mathrm{k}) &= - \rho_0\vec{g}(\vec{r}), \\
    \frac{\partial}{\partial t}\left(\rho_0 \frac{V^2}{2}+\frac{p}{\gamma
-1}+\frac{B_0^2}{2\mu}+\sum W_\mathrm{A,k}^\pm\right)+\nabla\cdot\left(\left[\rho_0\frac{V^2}{2}+\frac{p}{\gamma
-1}+\frac{B_0^2}{2\mu}\right]\vec{V}-\vec{B}_0\frac{\vec{V}\cdot\vec{B_0}}{\mu}\right) & \nonumber \\ +
\nabla\cdot\left(\vec{Z}_0^-W^+_\mathrm{A}+\vec{Z}_0^+W^-_\mathrm{A}+\vec{Q}_0^-W^+_\mathrm
{k}+\vec{Q}_0^+W^-_\mathrm{k}\right)+\nabla\cdot\left(\left[P_\mathrm{A}+
P_\mathrm{k}\right]\vec{V}\right)  =-{\cal
L} - \rho_0\vec{V}\cdot\vec{g}(\vec{r}) & + \frac{\zeta-1}{\zeta+1}P_\mathrm{k}\nabla\cdot\vec{V}, \label{eq:systemenergy} \\
    \frac{\partial W_\mathrm{A}^\pm}{\partial t} + \nabla
\cdot(\vec{Z}^\mp_0 W_\mathrm{A}^\pm) +  \frac{W_\mathrm{A}^\pm}{2} \nabla\cdot \vec{V} =
 - \Gamma^\mp W_\mathrm{A}^\pm \mp {\cal R}_\mathrm{A}, & \label{eq:alfven} \\
    \frac{\partial W_\mathrm{k}^\pm}{\partial t} + \nabla
\cdot(\vec{Q}^\mp_0W_\mathrm{k}^\pm) +  \frac{W_\mathrm{k}^\pm}{2} \nabla\cdot \vec{V} =
 - \frac{1}{L_{\perp,\mathrm{VD}}} \frac{1}{\sqrt{\rho_\mathrm{e}}}
(W_\mathrm{k}^\pm)^{3/2} \mp {\cal R}_\mathrm{k} . & \label{eq:kink}
\end{align}

In all of these equations, $\rho_0$ is actually $\rho_\mathrm{e}$. When averaging the density over a cross-section, we have
\[\frac{1}{\pi \gamma^2 R^2} \int d\phi \int_0^{\gamma R} \rho rdr = \frac{1}{\pi \gamma^2 R^2} \left( \pi R^2 \rho_\mathrm{i} + \pi (\gamma^2
R^2-R^2) \rho_\mathrm{e}\right)=f\rho_\mathrm{i} + (1-f)\rho_\mathrm{e}=\rho_\mathrm{e}+f(\rho_\mathrm{i}-\rho_\mathrm{e}).\]
This averaging over the cross-section of the density would occur in the continuity equation and the non-wave pressure terms of the momentum equation.
This understanding that the modelled density $\rho_0$ is really the external density $\rho_\mathrm{e}$ has no impact on the modelling that is
performed (see next section). However, appropriate values must be taken. Moreover, additional care is needed when implementing radiative losses, or
doing forward modelling, both of which are non-linearly weighed with the density.

In Eqs.~\ref{eq:alfven}-\ref{eq:kink} we have included ad-hoc reflection terms. It is important that these reflection terms adhere to the basic 
principles of conservation of energy: what is taken out of the $+$-equation must be injected as a source term in the $-$-equation, and vice versa. 
Several expressions for the reflection of \alfven{} waves exist in the literature \citep[e.g.][]{vanderholst2014,downs2016,reville2020}. It is less 
clear what the reflection term for kink waves might be. Surely, it must be proportional to the strength of the longitudinal kink speed gradient. 
Inspiration for this aspect may be found also in the existence of a cut-off period for kink waves \citep{pelouze2023}. 

After a lot of calculations, it seems that the evolution equation for \alfven{} and kink waves (Eq.~\ref{eq:alfven} and Eq.~\ref{eq:kink} 
respectively) are eerily similar. However, there is a key difference. On the one hand, the damping term for the \alfven{} waves is dependent on the 
amplitude of the counterpropagating wave \citep[as has been known for a long time, at least in the homogeneous or WKB picture, while in a stratified 
plasma, the \alfven{} waves also acquire a mixed character enabling cascade][]{velli1989,verdini2009}. Thus, upward-propagating \alfven{} waves in 
plumes 
are hardly damping in the low solar atmosphere because insufficient reflected waves have been generated. On the other hand, the damping term for the 
kink wave due to uniturbulence only depends on the amplitude of the kink wave itself (and not on its counterpropagating wave). Thus, the damping is 
at work, even if no counterpropagating waves are generated yet. This is of particular importance in the low part of the solar atmosphere. There the 
kink waves damp immediately and significantly \citep[damping lengths that are between 0.25 and 10$\times$ their wavelengths, i.e. 
25Mm-4000Mm,][]{vd2020}, providing a crucial ingredient in kick-starting the solar wind and coronal heating low down. We thus expect to have a major 
contribution of the kink wave heating in the low corona, whereas the traditional \alfven{} wave heating takes over higher up (say around $1R_\sun$). 

A limitation to these wave equations (for the \alfven{} wave as well as the kink wave) is that it has been derived for in the WKB limit. We 
\citep[but also previous authors][]{vanderholst2014,reville2020} have thus used the implicit assumption that the background variables are only slowly 
varying along the magnetic field. This allows the waves to be described by only a single \elsasser{} or $Q$-variable, and decouples the wave modes. 
However, if the stratification is significant, the waves are modified and will have a mixed character \citep{velli1989,verdini2009}. Thus, the 
full, true evolution equations will be much more complicated and retain their 3D nature \citep{zank2012}. 

\section{Numerical implementation}\label{sec:numerics}
To show a first proof-of-concept, we have implemented the UAWSOM equations (Eqs.~\ref{eq:continuity}-\ref{eq:kink}) in 1D in python using the {\em 
fipy} package\footnote{\url{https://www.ctcms.nist.gov/fipy/}} \citep{fipy}. However, for this proof-of-concept, we have neglected the \alfven{} wave 
equation (Eq.~\ref{eq:alfven}, corresponding to taking $W_\mathrm{A}^\pm\equiv 0$ in the entire simulation). For ease of
implementation, we use the equation for internal energy (Eq.~\ref{eq:gasp}) instead of Eq.~\ref{eq:systemenergy}. In 1D, the system of equations
(Eqs.~\ref{eq:continuity}-\ref{eq:kink}) is reduced to a coupled set of 5 differential equations, since the velocity only has a component along $z$ 
and the induction equation is reduced to $B(z)$ constant (as a function of time). The resulting python implementation serves as a reference
implementation for
the UAWSOM concept, and has been made available 
on Gitlab\footnote{\url{https://gitlab.kuleuven.be/plasma-astrophysics/research/tom_s-projects/uawsom.git}}. In this paper, we have used revision
{\tt 95111d80}. In order to check if the wave propagation equations are correctly implemented, we have also performed a simplified test in a
stationary background. The python script for that test is also accessible through the Gitlab repository.

In the equations (Eqs.~\ref{eq:continuity}-\ref{eq:kink}), we have incorporated the radiative losses, by using the implementation of 
\citet{hermans2021} of the coronal cooling curve of
\citet{dere2009}. We interpolate the cooling curve's tabulated points with a cubic spline to the temperatures needed in the simulation. On top of the
terms included in the energy equation above (Eq.~\ref{eq:gasp}), we have also incorporated a thermal conduction term as part of its RHS:
\begin{equation}
    \mathrm{Thermal\ conduction} = - \nabla \cdot (\kappa \nabla T),
\end{equation}
with $\kappa= \kappa_0 T^{5/2}$ and $\kappa_0 = 8\ 10^{-7} \mathrm{erg}/\mathrm{cm}/\mathrm{s}/\mathrm{K}^{7/2}$. In this first implementation, we
have neglected the reflection terms ${\cal R}_\mathrm{k}=0$ in the equation for the kink wave energy (Eq.~\ref{eq:kink}). The correct values are not
known \citep{pelouze2023} and should be investigated in a separate modelling paper.
Gravity
is fixed at $\vec{g}(z)=274\mathrm{m}/\mathrm{s}^2 \vec{1}_z$.\\

The python programme has as option to read in an {\em numpy} savefile for its initial condition. However, here we show some results from its start of
a pre-implemented plasma configuration. In that pre-implemented configuration, we use as equilibrium 
\begin{align}
    \rho(z) & = \rho_0 \exp{(-z/H)}, \label{eq:background_rho}\\
    V(z) & = 0, \label{eq:V_of_height}\\
    B(z) & = B_0 \frac{R_\sun^2}{\left(z+R_\sun\right)^2},\label{eq:background_B}\\
    R(z) & = R_0 \frac{B_0^2}{B(z)^2},\label{eq:background_R}\\
    \zeta & = (\zeta_0-1)\exp{\left(-\frac{z}{5R_\sun}\right)}+1, \label{eq:background_zeta}
\end{align}
with the initial values $\rho_0$ is taken equivalent to a number density of $10^9\mathrm{cm}^{-3}$, $H=50\mathrm{Mm}$, 
$B_0=20\mathrm{G}$, $\zeta_0=5$, and $R_0=1\mathrm{Mm}$. We fix the
initial temperature in the entire domain at $T_0=1\mathrm{MK}$.

The initial conditions for $W^\pm_\mathrm{k}$ are taken as
\begin{align}
    W^-_\mathrm{k}&=W_0\left(\frac{1}{2}\cos{\left(\frac{\pi z}{\delta}\right)}+\frac{1}{2}\right), &\mbox{for }
z<\delta,\\
    W^-_\mathrm{k}&=0, & \mbox{elsewhere},\\
    W^+_\mathrm{k}&=0,
\end{align}
with $\delta = 30\mathrm{Mm}$ and $W_0=4\ 10^{-4} \mathrm{J}/\mathrm{m}^3$. This smooth initial condition for $W_\mathrm{k}^-$ was chosen so that an
initial abrupt start from a non-zero boundary condition would not lead to difficulties with the boundary conditions.

\begin{figure}[tb]
    \includegraphics[width=.9\linewidth]{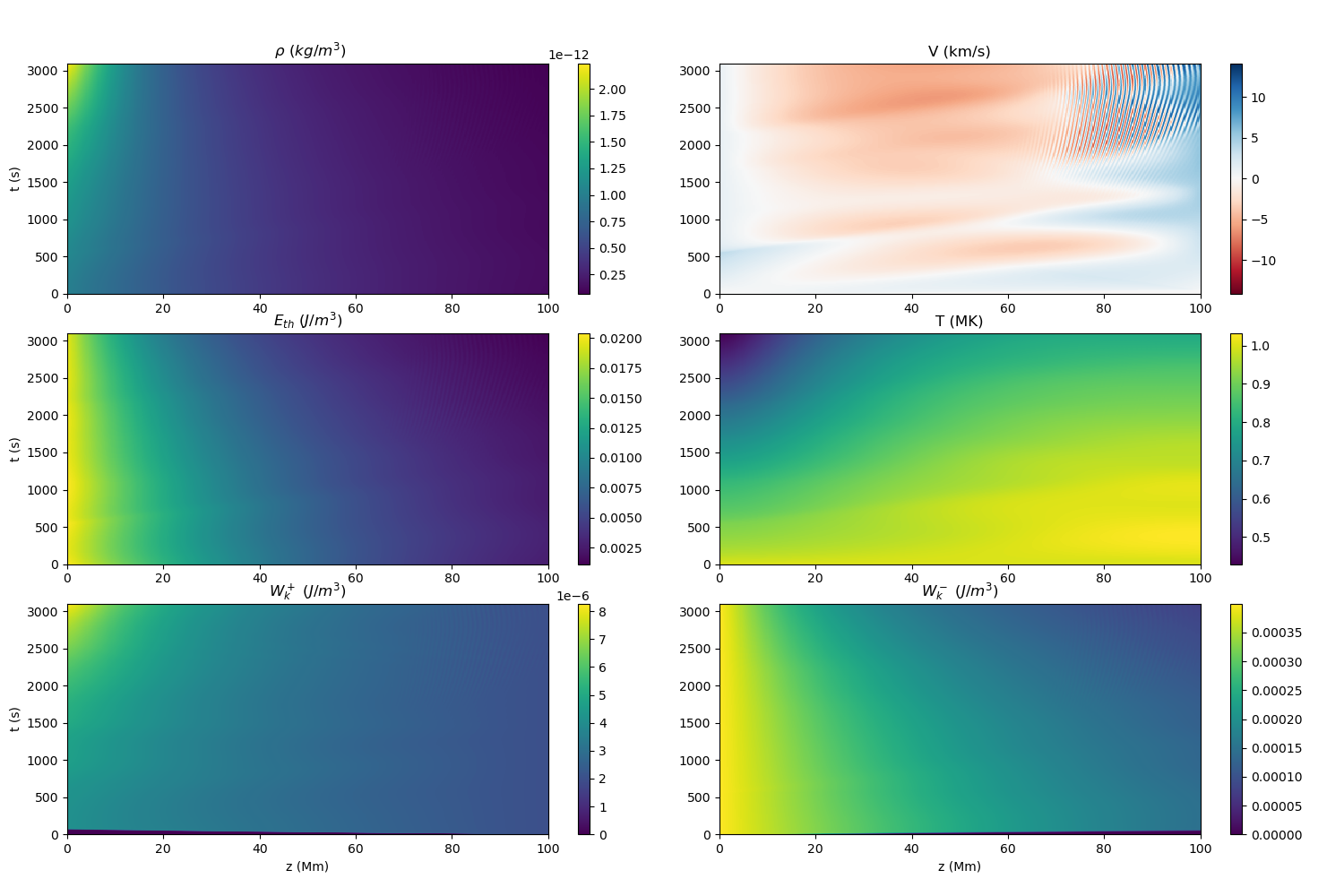}
    \caption{The continuation of the simulation with kink wave heating and radiative losses until a time of $3100\mathrm{s}$. }
    \label{fig:longsim}
\end{figure}

As boundary conditions we take the following at the bottom boundary:
\begin{align}
    \frac{d\rho}{dz} &= 0,\\
    V & = 0, \\
    \frac{dE_\mathrm{th}}{dz}= \frac{1}{\gamma-1}\frac{dp}{dz} & = -\frac{g\rho}{\gamma-1}, \\
    W_\mathrm{k}^- & = W_0, \\
    \frac{dW_\mathrm{k}^+}{dz} & = 0,
\end{align}
where the condition for the thermal energy $E_\mathrm{th}$ expresses a continuous isothermal stratification beyond the computational domain. At the 
top
boundary, we put as boundary conditions
\begin{align}
    \frac{d\rho}{dz} &= 0,\\
    \frac{dV}{dz} & = 0, \\
    \frac{dE_\mathrm{th}}{dz}= \frac{1}{\gamma-1}\frac{dp}{dz} & = -\frac{g\rho}{\gamma-1}, \\
    \frac{dW_\mathrm{k}^-}{dz} & = 0,\\
    W_\mathrm{k}^+ & = \frac{W_0}{200}.
\end{align}
Our aim was to have reasonably open boundary conditions at the top. However, Fig.~\ref{fig:longsim} shows that there are still reflected 
slow waves in
the simulation as can be seen from bouncing flows between the boundaries.

\begin{figure}[tb]
    \includegraphics[width=.9\linewidth]{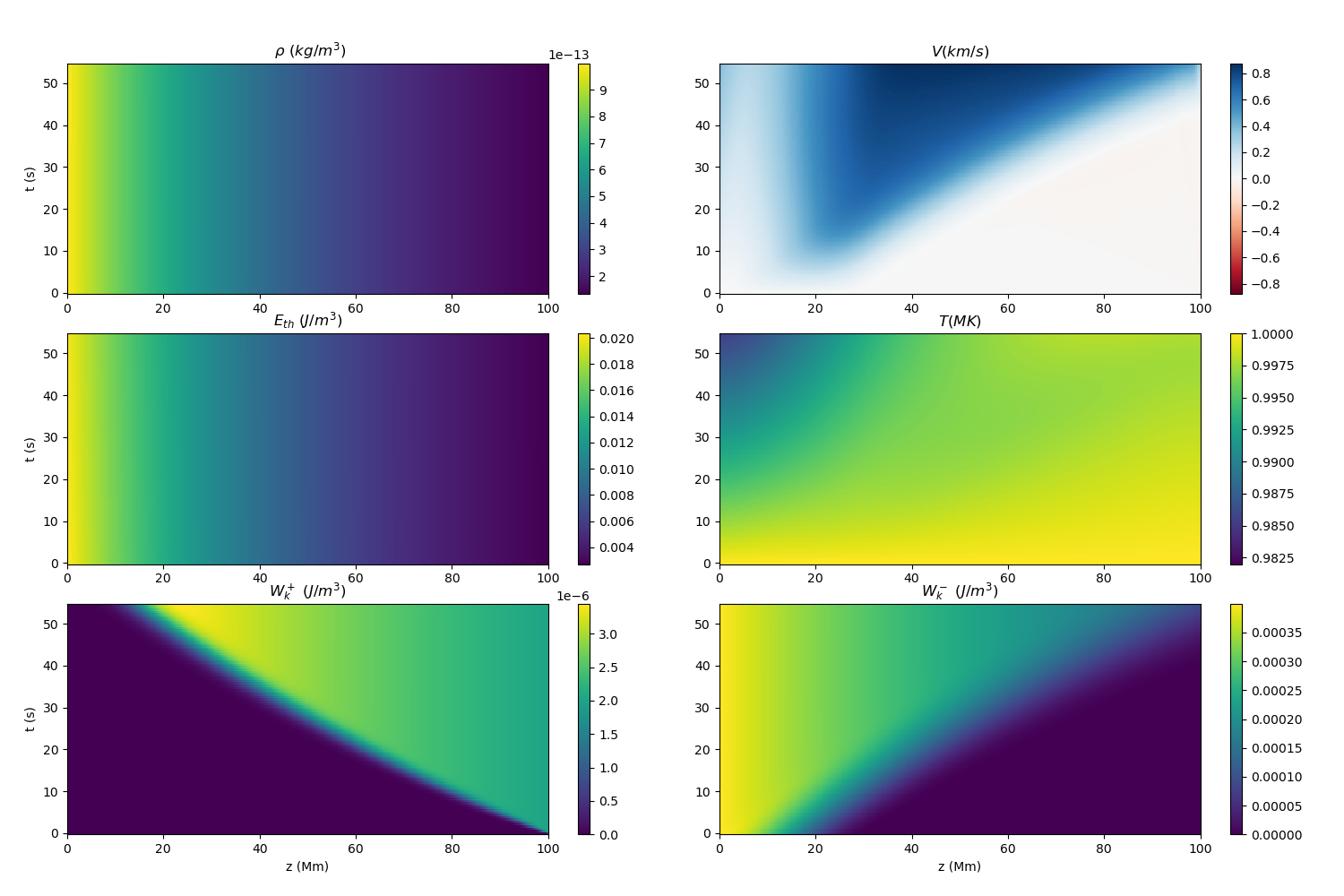}
    \includegraphics[width=.9\linewidth]{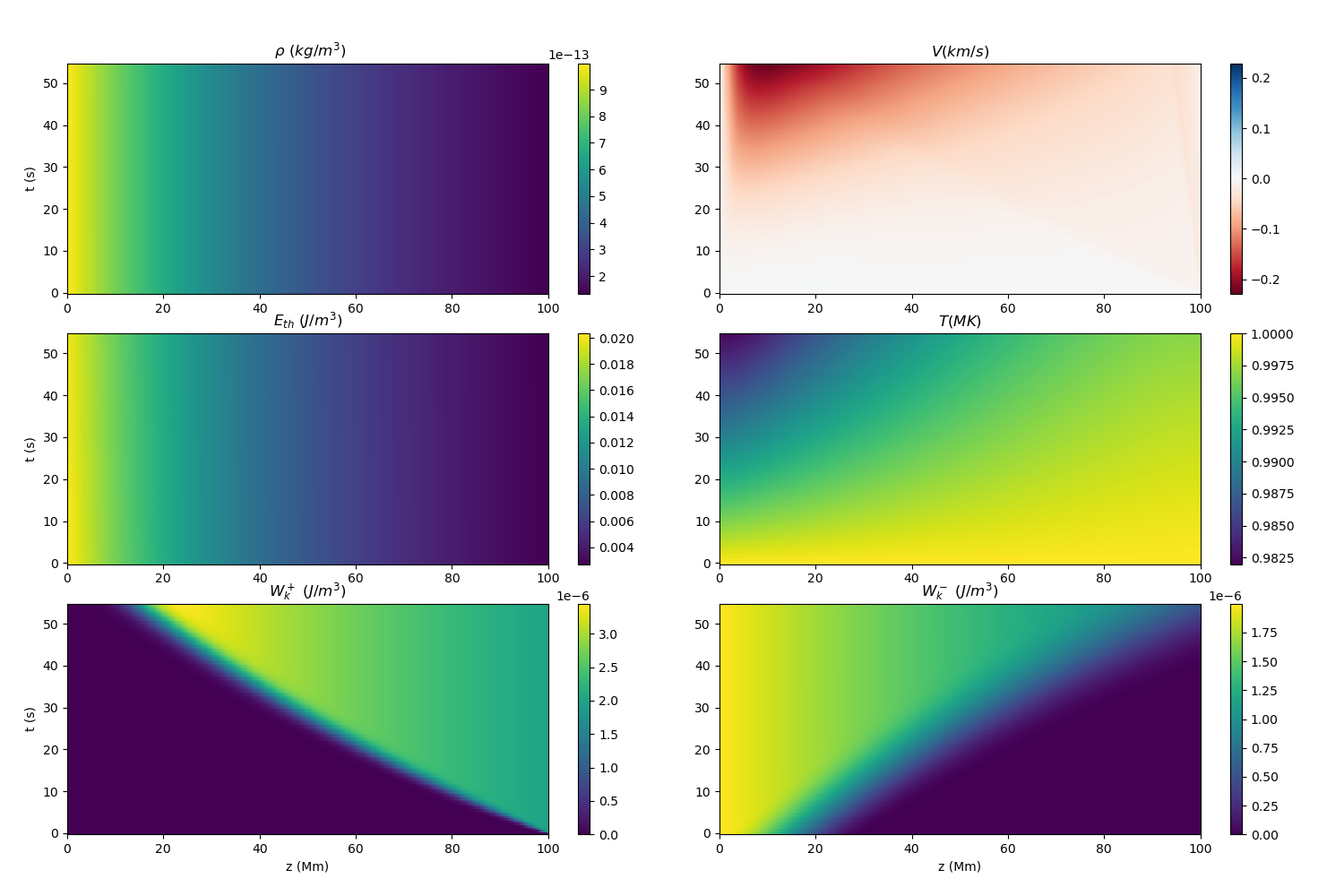}
    \caption{The first 1000 time steps for the large simulation. Top 6 panels: Simulation variables for the simulation with wave heating. As in the 
following figures,
the 6 panels consist of density $\rho$, velocity $V$, thermal energy $E_\mathrm{th}$, temperature $T$ and wave energies $W_\mathrm{k}^\pm$. Bottom 6
panels: The same, but with a much lower influx value of $W_\mathrm{k}^-=W_0/200$ at the left boundary, resulting in almost no kink wave
heating.}
    \label{fig:firststeps}
\end{figure}

\begin{figure}[tb]
    \centering
    \includegraphics[width=.7\linewidth]{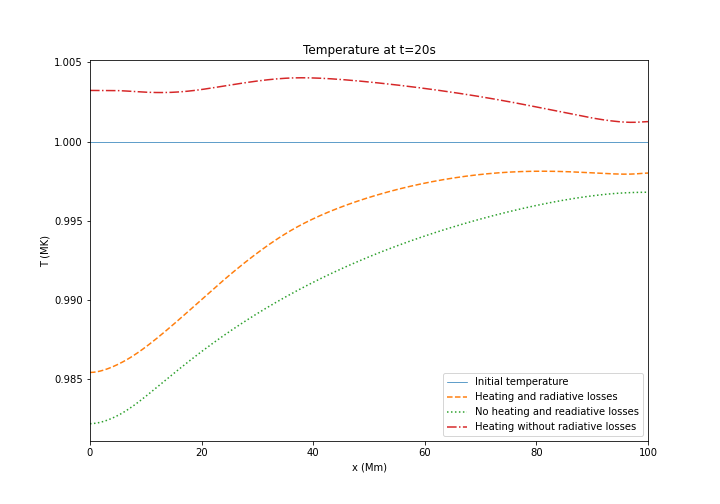}
    \caption{The temperature profile as a function of height at the end of the short simulations shown in Fig.~\ref{fig:firststeps}, for 3 different 
simulations: one with kink wave heating and no radiative losses, one with kink wave heating and radiative losses, and one without kink wave heating 
and radiative losses.}
    \label{fig:temp}
\end{figure}

We display the first $55\mathrm{s}$ of runtime (1000 iteration steps with {\em fipy}) in Fig.~\ref{fig:firststeps}. The top 6 panels in the figure
show the system variables ($\rho$, $V$, $E_\mathrm{th}$, $W_\mathrm{k}^\pm$) and also the temperature $T$. The bottom 6 panels show the simulation
but with a much lower energy $W_\mathrm{k}^-$ in the kink wave (by setting the left boundary condition for this variable to $W_0/200$). As visible in
the bottom right panel (of the top 6), the kink wave is propagating into the system with the appropriate kink speed. The resulting energy dissipation
and deposition is leading to heating, as can be seen by comparing the temperature scale of the $T$-panel of the top 6 and the bottom 6. The
temperature $T$ is gradually decreasing in both, because of the radiative losses. However, the top panel with the kink wave heating has a slightly
higher temperature than the non-heated simulation. This is more clearly shown in Fig.~\ref{fig:temp}, where the final temperature profile as a 
function of height is shown for the top and bottom simulation of Fig.~\ref{fig:firststeps}, corresponding to the case of ``heating'' and ``no 
heating''. Additionally, we display the initial temperature of 1MK and also a simulation with kink wave heating, but where the radiative losses have 
been switched off. The graph in Fig.~\ref{fig:temp} clearly shows that the wave heating indeed increases the temperature, either above the initial 
temperature in the case of no radiative losses, or at least above the temperature in a simulation with only just radiative cooling.\par

The effect of the kink wave pressure $P_\mathrm{k}$ is also visible in the panels of Fig.~\ref{fig:firststeps} 
showing the velocity $V$. The top right panel with kink wave pressure shows a gentle upflow, while the bottom panels show a gradual downflow. The
latter happens because of the restructuring of the plasma due to the radiative cooling. Also visible in that panel ($V$ of the bottom 6 panels) is the
effect of the top boundary, which has a propagating signal associated with the downward kink wave, but also a signal propagating with the sound
speed. \\
The effect of the numerical diffusivity is shown in the bottom left panel $W_\mathrm{k}^+$. The front of the 
$W_\mathrm{k}^+$ energy should
theoretically remain a sharp interface, but it is clear that during its propagation it is smoothed out.

In Fig.~\ref{fig:longsim}, we show the longer duration run of the simulation with the kink wave heating, for $3100\mathrm{s}$. In the panel with 
velocity $V$, it is shown that
there is a gradual upflow generated by the kink wave driving, on the order of $5-10\mathrm{km}/\mathrm{s}$. However, at times later than
$2000\mathrm{s}$ and height above $70\mathrm{Mm}$, numerical instabilities show up, which eventually crash the simulation. Some
additional viscosity terms might damp out the instabilities. Still, the panel clearly shows that upflows are driven by the kink wave pressure,
because these upflows are absent in the non-driven simulation. \\
As can be seen in the temperature panel $T$, the temperature gradually decreases. However, it does not decrease as much as expected without wave
heating. The cooling times in the simulation, calculated as $\tau = E_\mathrm{th}/{\cal L}$ are between $2700-20000\mathrm{s}$. In the panel, it is
clearly visible that the plasma is not cooling down as much as expected from these cooling time scales, given that the temperature's minimum value is
$0.45\mathrm{MK}> T_0/e$. Thus, the kink waves are indeed partially heating the plasma, or at least sustaining it against radiative losses
\citep[similar to][]{shi2021}.\\
The cooling by the radiative losses is not entirely counteracted, and this is shown by the strong decrease of the initial temperature. As a result of
the steady cooling, the plasma is restructuring to fit with the new quasi-equilibrium. That leads to a drainage of density towards the footpoint, as
is visible in the top left panel.\par

It is clear that this simulation is not the final simulation needed for concluding the UAWSOM model and determining its usefulness. Further work is
needed, particularly for varying equilibrium parameters and driving parameters. It was not our aim to obtain a steady solution for the solar
atmosphere, but rather a proof-of-concept for the application of the UAWSOM equations (Eqs.~\ref{eq:continuity}-\ref{eq:kink}).


\section{Conclusions}
In this paper, we have employed the formalism of $Q$-variables. This formalism gives equations co-propagating with a specific wave. We have used the
equations of \citet{vd2024} and recast the equations to an energy propagation equation. We have found that all waves show self-damping through
non-linear effects, except for \alfven{} waves. This is visible in the energy equation (Eq.~\ref{eq:wqevol}) because of terms proportional to $\Delta
\alpha^2$, a parameter which is always non-zero except for \alfven{} waves.\par

We have further specialised the energy propagation equation to the case of kink waves. To that end, we have followed the approach of \citet{vd2020}
and averaged out the wave energy equation over the cross-section of the plume. From that averaging process, we have found a single equation for kink
wave energy evolution (Eq.~\ref{eq:unidiss}). This equation introduced the $L_{\perp,\mathrm{VD}}$ factor, but unlike \alfven{} wave turbulence this 
parameter is
given in terms of background properties. We have found that the energy dissipation term is proportional to $(W_\mathrm{k}^\pm)^{3/2}$. \\
Such a proportionality of the damping term was also previously proposed by \citet{zank1996}.  However, in his case, the damping term was multiplied
with the $f(\sigma_c)$ (which becomes zero if cross-helicity $\sigma_c=\pm 1$, i.e. only one of the counterpropagating waves). Here, on the other
hand, the $f(\sigma_c)$ remains non-zero even if only one of the propagation directions is present. Such a case was observationally considered by
\citet{adhikari2023}. The presence of this term in the equation shows that kink waves (or any other wave, except for \alfven{} waves) will damp even
if a wave is propagating in only one direction occurs.\par

Subsequently, we have combined the energy propagation equation for kink waves (Eq.~\ref{eq:unidiss}), with a slowly varying (in the
WKB sense) background that is governed by the MHD equations. The resulting system (Eqs.~\ref{eq:continuity}-\ref{eq:kink}, which we have dubbed the
UAWSOM system) has the traditional conservation of mass, momentum and energy, but is supplemented by wave energy equations. The latter wave energy
equations keep track of the wave energy and the wave's damping. The wave energy is then added to the total energy and incorporated in the energy
conservation equation. This ensures that energy loss of the wave results in the increase of thermal energy and thus heating of the plasma.\\
For now, we have considered this system of equations to be an extension to the classical AWSOM model of \citet{vanderholst2014}. On top of the AWSOM
model, we have additionally incorporated the kink wave driving. The kink wave evolution equation is apparently very similar to the \alfven{} wave 
evolution equation. However, the difference is in the damping term. The \alfven{} waves require counterpropagating waves to initiate the cascade to 
turbulent heating and low in the atmosphere the reflected waves are not strong enough to generate sufficient heating. The damping term for the kink 
waves, however, is also operating when there is no counterpropagating wave. Thus, the kink waves can already start heating the plasma low down in the 
atmosphere where the damping is further enhanced by the large density gradients of the flux tubes (and despite the absence of reflected waves). It 
may thus well be that kink waves play a key role in providing the jump start to solar coronal and solar wind heating low down in (say) the first few 
100Mm. 
\\
In principle, the formalism for the addition of a kink wave equation can also be used to incorporate a 3rd equation on the
slow wave driving and development. That would allow us to model
the interaction of the parametric instability and its influence on the reflection of the \alfven{} waves. Then, we would be able to capture the
essential physics in the \citet{shoda2019} paper in a similar 1D model. This is left for a future work.\par

Lastly, we have implemented the UAWSOM equations in a python module, where we considered the 1D version of the 
full set of equations (Eqs.~\ref{eq:continuity}-\ref{eq:kink}, albeit with \alfven{} wave power set to 0). We have shown that the kink wave driving 
of the solar atmosphere
leads to the generation of upflows. Moreover, the uniturbulent damping of the kink waves leads to the heating of the plasma, which seems most 
significant in the low corona. In some regions in our
simulation, the kink wave heating could sustain the plasma against the cooling by radiative losses, but in others it could not. These initial results 
with the python
implementation show that further work is needed, but it also showcases the potential that the UAWSOM model holds. As such, the implementation offered
in this article serves as a proof-of-concept rather than a full investigation of this model's usefulness. \\ As future work, we will compare 1D but 
also 3D solar atmospheres which are driven solely by \alfven{} waves, solely by kink waves and mixed models (with varying ratios of kink and 
\alfven{} wave energy content). The comparison of these models to observations will allow us to place bounds on the ratio of kink wave and \alfven{} 
wave driving energy.

\begin{acknowledgements}
      TVD was supported by the European Research Council (ERC) under the European Union's Horizon 2020 research and innovation programme (grant 
agreement No 724326), the C1 grant TRACEspace of Internal Funds KU Leuven, and a Senior Research Project (G088021N) of the FWO Vlaanderen.
Furthermore, TVD received financial support from the Flemish Government under the long-term structural Methusalem funding program, project SOUL:
Stellar evolution in full glory, grant METH/24/012 at KU Leuven. The research that led to these results was subsidised by the Belgian Federal Science
Policy Office through the contract B2/223/P1/CLOSE-UP. It is also part of the DynaSun project and has thus received funding under the Horizon
Europe programme of the European Union under grant agreement (no. 101131534). Views and opinions expressed are however those of the author(s) only and
do not necessarily reflect those of the European Union and therefore the European Union cannot be held responsible for them. TVD is grateful for the
hospitality of I\~nigo Arregui (IAC, Tenerife) where these results were discussed.\\
NM acknowledges Research Foundation – Flanders (FWO Vlaanderen) for their support through a Postdoctoral Fellowship.\\
MVS acknowledge support from the French Research Agency grant ANR STORMGENESIS \#ANR-22-CE31-0013-01.\\
This research was supported by the International Space Science Institute (ISSI) in Bern, through ISSI International Team project \#560.
\end{acknowledgements}

%
\bibliographystyle{aa} 
\bibliography{refs} 

\begin{thebibliography}{68}
\expandafter\ifx\csname natexlab\endcsname\relax\def\natexlab#1{#1}\fi

\bibitem[{{Adhikari} {et~al.}(2023){Adhikari}, {Zank}, {Telloni}, {Zhao}, \&
  {Pitna}}]{adhikari2023}
{Adhikari}, L., {Zank}, G.~P., {Telloni}, D., {Zhao}, L.~L., \& {Pitna}, A.
  2023, in Journal of Physics Conference Series, Vol. 2544, Journal of Physics
  Conference Series (IOP), 012007

\bibitem[{{Alvarado-G{\'o}mez} {et~al.}(2022){Alvarado-G{\'o}mez}, {Cohen},
  {Drake}, {Fraschetti}, {Poppenhaeger}, {Garraffo}, {Chebly}, {Ilin},
  {Harbach}, \& {Kochukhov}}]{alvarado2022}
{Alvarado-G{\'o}mez}, J.~D., {Cohen}, O., {Drake}, J.~J., {et~al.} 2022, \apj,
  928, 147

\bibitem[{{Anfinogentov} {et~al.}(2013){Anfinogentov}, {Nakariakov},
  {Mathioudakis}, {Van Doorsselaere}, \& {Kowalski}}]{anfinogentov2013}
{Anfinogentov}, S., {Nakariakov}, V.~M., {Mathioudakis}, M., {Van
  Doorsselaere}, T., \& {Kowalski}, A.~F. 2013, \apj, 773, 156

\bibitem[{{Anfinogentov} {et~al.}(2015){Anfinogentov}, {Nakariakov}, \&
  {Nistic{\`o}}}]{anfinogentov2015}
{Anfinogentov}, S.~A., {Nakariakov}, V.~M., \& {Nistic{\`o}}, G. 2015, \aap,
  583, A136

\bibitem[{{Antolin} \& {Van Doorsselaere}(2019)}]{antolin2019}
{Antolin}, P. \& {Van Doorsselaere}, T. 2019, Frontiers in Physics, 7, 85

\bibitem[{{Banerjee} {et~al.}(1998){Banerjee}, {Teriaca}, {Doyle}, \&
  {Wilhelm}}]{banerjee1998}
{Banerjee}, D., {Teriaca}, L., {Doyle}, J.~G., \& {Wilhelm}, K. 1998, \aap,
  339, 208

\bibitem[{{Bruno} \& {Carbone}(2013)}]{bruno2013}
{Bruno}, R. \& {Carbone}, V. 2013, Living Reviews in Solar Physics, 10, 2

\bibitem[{{Cohen} {et~al.}(2024){Cohen}, {Glocer}, {Garraffo},
  {Alvarado-G{\'o}mez}, {Drake}, {Monsch}, \& {Fauth Puigdomenech}}]{cohen2024}
{Cohen}, O., {Glocer}, A., {Garraffo}, C., {et~al.} 2024, \apj, 962, 157

\bibitem[{{De Moortel} \& {Howson}(2022)}]{demoortel2022}
{De Moortel}, I. \& {Howson}, T.~A. 2022, \apj, 941, 85

\bibitem[{{De Pontieu} {et~al.}(2012){De Pontieu}, {Carlsson}, {Rouppe van der
  Voort}, {Rutten}, {Hansteen}, \& {Watanabe}}]{depontieu2012}
{De Pontieu}, B., {Carlsson}, M., {Rouppe van der Voort}, L.~H.~M., {et~al.}
  2012, \apjl, 752, L12

\bibitem[{{Dere} {et~al.}(2009){Dere}, {Landi}, {Young}, {Del Zanna},
  {Landini}, \& {Mason}}]{dere2009}
{Dere}, K.~P., {Landi}, E., {Young}, P.~R., {et~al.} 2009, \aap, 498, 915

\bibitem[{{Downs} {et~al.}(2016){Downs}, {Lionello}, {Miki{\'c}}, {Linker}, \&
  {Velli}}]{downs2016}
{Downs}, C., {Lionello}, R., {Miki{\'c}}, Z., {Linker}, J.~A., \& {Velli}, M.
  2016, \apj, 832, 180

\bibitem[{{Evans} {et~al.}(2012){Evans}, {Opher}, {Oran}, {van der Holst},
  {Sokolov}, {Frazin}, {Gombosi}, \& {V{\'a}squez}}]{evans2012}
{Evans}, R.~M., {Opher}, M., {Oran}, R., {et~al.} 2012, \apj, 756, 155

\bibitem[{{Evensberget} \& {Vidotto}(2024)}]{evensberget2024}
{Evensberget}, D. \& {Vidotto}, A.~A. 2024, \mnras, 529, L140

\bibitem[{{Gao} {et~al.}(2022){Gao}, {Tian}, {Van Doorsselaere}, \&
  {Chen}}]{gao2022}
{Gao}, Y., {Tian}, H., {Van Doorsselaere}, T., \& {Chen}, Y. 2022, \apj, 930,
  55

\bibitem[{{Goossens} {et~al.}(2012){Goossens}, {Andries}, {Soler}, {Van
  Doorsselaere}, {Arregui}, \& {Terradas}}]{goossens2012}
{Goossens}, M., {Andries}, J., {Soler}, R., {et~al.} 2012, \apj, 753, 111

\bibitem[{{Goossens} {et~al.}(1992){Goossens}, {Hollweg}, \&
  {Sakurai}}]{goossens1992}
{Goossens}, M., {Hollweg}, J.~V., \& {Sakurai}, T. 1992, \solphys, 138, 233

\bibitem[{{Goossens} {et~al.}(2009){Goossens}, {Terradas}, {Andries},
  {Arregui}, \& {Ballester}}]{goossens2009}
{Goossens}, M., {Terradas}, J., {Andries}, J., {Arregui}, I., \& {Ballester},
  J.~L. 2009, \aap, 503, 213

\bibitem[{{Goossens} {et~al.}(2013){Goossens}, {Van Doorsselaere}, {Soler}, \&
  {Verth}}]{goossens2013}
{Goossens}, M., {Van Doorsselaere}, T., {Soler}, R., \& {Verth}, G. 2013, \apj,
  768, 191

\bibitem[{{Guo} {et~al.}(2019){Guo}, {Van Doorsselaere}, {Karampelas}, {Li},
  {Antolin}, \& {De Moortel}}]{guo2019b}
{Guo}, M., {Van Doorsselaere}, T., {Karampelas}, K., {et~al.} 2019, \apj, 870,
  55

\bibitem[{{Guyer} {et~al.}(2009){Guyer}, {Wheeler}, \& {Warren}}]{fipy}
{Guyer}, J.~E., {Wheeler}, D., \& {Warren}, J.~A. 2009, Computing in Science
  and Engineering, 11, 6

\bibitem[{{Hahn} \& {Savin}(2013)}]{hahn2013}
{Hahn}, M. \& {Savin}, D.~W. 2013, \apj, 776, 78

\bibitem[{{Hermans} \& {Keppens}(2021)}]{hermans2021}
{Hermans}, J. \& {Keppens}, R. 2021, \aap, 655, A36

\bibitem[{{Hillier} {et~al.}(2019){Hillier}, {Barker}, {Arregui}, \&
  {Latter}}]{hillier2019}
{Hillier}, A., {Barker}, A., {Arregui}, I., \& {Latter}, H. 2019, \mnras, 482,
  1143

\bibitem[{{Ismayilli} {et~al.}(2022){Ismayilli}, {Van Doorsselaere},
  {Goossens}, \& {Magyar}}]{ismayilli2022}
{Ismayilli}, R., {Van Doorsselaere}, T., {Goossens}, M., \& {Magyar}, N. 2022,
  Frontiers in Astronomy and Space Sciences, 8, 241

\bibitem[{{Ismayilli} {et~al.}(2024){Ismayilli}, {Van Doorsselaere}, {Magyar},
  {Changmai}, \& {Verdini}}]{ismayilli2024}
{Ismayilli}, R., {Van Doorsselaere}, T., {Magyar}, N., {Changmai}, M., \&
  {Verdini}, A. 2024, Physics of Fluids, 36, 065109

\bibitem[{{Karampelas} \& {Van Doorsselaere}(2018)}]{karampelas2018}
{Karampelas}, K. \& {Van Doorsselaere}, T. 2018, \aap, 610, L9

\bibitem[{{Karampelas} \& {Van Doorsselaere}(2021)}]{karampelas2021}
{Karampelas}, K. \& {Van Doorsselaere}, T. 2021, \apjl, 908, L7

\bibitem[{{Lim} {et~al.}(2023){Lim}, {Van Doorsselaere}, {Berghmans}, {Morton},
  {Pant}, \& {Mandal}}]{lim2023}
{Lim}, D., {Van Doorsselaere}, T., {Berghmans}, D., {et~al.} 2023, \apjl, 952,
  L15

\bibitem[{{Magyar} \& {Nakariakov}(2021)}]{magyar2021}
{Magyar}, N. \& {Nakariakov}, V.~M. 2021, \apj, 907, 55

\bibitem[{Magyar {et~al.}({2017})Magyar, Van~Doorsselaere, \&
  Goossens}]{magyar2017}
Magyar, N., Van~Doorsselaere, T., \& Goossens, M. {2017}, {Nat. Sci. Rep.}, {7}

\bibitem[{{Magyar} {et~al.}(2019){Magyar}, {Van Doorsselaere}, \&
  {Goossens}}]{magyar2019}
{Magyar}, N., {Van Doorsselaere}, T., \& {Goossens}, M. 2019, \apj, 882, 50

\bibitem[{{Morton} {et~al.}(2021){Morton}, {Tiwari}, {Van Doorsselaere}, \&
  {McLaughlin}}]{morton2021}
{Morton}, R.~J., {Tiwari}, A.~K., {Van Doorsselaere}, T., \& {McLaughlin},
  J.~A. 2021, \apj, 923, 225

\bibitem[{{Nakariakov} {et~al.}(2024){Nakariakov}, {Zhong}, {Kolotkov},
  {Meadowcroft}, {Zhong}, \& {Yuan}}]{nakariakov2024}
{Nakariakov}, V.~M., {Zhong}, S., {Kolotkov}, D.~Y., {et~al.} 2024, Reviews of
  Modern Plasma Physics, 8, 19

\bibitem[{{Nechaeva} {et~al.}(2019){Nechaeva}, {Zimovets}, {Nakariakov}, \&
  {Goddard}}]{nechaeva2019}
{Nechaeva}, A., {Zimovets}, I.~V., {Nakariakov}, V.~M., \& {Goddard}, C.~R.
  2019, \apjs, 241, 31

\bibitem[{{Nistic{\`o}} {et~al.}(2013){Nistic{\`o}}, {Nakariakov}, \&
  {Verwichte}}]{nistico2013}
{Nistic{\`o}}, G., {Nakariakov}, V.~M., \& {Verwichte}, E. 2013, \aap, 552, A57

\bibitem[{{Pant} \& {Van Doorsselaere}(2020)}]{pant2020}
{Pant}, V. \& {Van Doorsselaere}, T. 2020, \apj, 899, 1

\bibitem[{{Parashar} {et~al.}(2020){Parashar}, {Goldstein}, {Maruca},
  {Matthaeus}, {Ruffolo}, {Bandyopadhyay}, {Chhiber}, {Chasapis}, {Qudsi},
  {Vech}, {Roberts}, {Bale}, {Bonnell}, {de Wit}, {Goetz}, {Harvey},
  {MacDowall}, {Malaspina}, {Pulupa}, {Kasper}, {Korreck}, {Case}, {Stevens},
  {Whittlesey}, {Larson}, {Livi}, {Velli}, \& {Raouafi}}]{parashar2020}
{Parashar}, T.~N., {Goldstein}, M.~L., {Maruca}, B.~A., {et~al.} 2020, \apjs,
  246, 58

\bibitem[{{Pelouze} {et~al.}(2023){Pelouze}, {Van Doorsselaere}, {Karampelas},
  {Riedl}, \& {Duckenfield}}]{pelouze2023}
{Pelouze}, G., {Van Doorsselaere}, T., {Karampelas}, K., {Riedl}, J.~M., \&
  {Duckenfield}, T. 2023, \aap, 672, A105

\bibitem[{{Petrova} {et~al.}(2023){Petrova}, {Magyar}, {Van Doorsselaere}, \&
  {Berghmans}}]{petrova2023}
{Petrova}, E., {Magyar}, N., {Van Doorsselaere}, T., \& {Berghmans}, D. 2023,
  \apj, 946, 36

\bibitem[{{Petrova} {et~al.}(2024){Petrova}, {Van Doorsselaere}, {Berghmans},
  {Parenti}, {Valori}, \& {Plowman}}]{petrova2024}
{Petrova}, E., {Van Doorsselaere}, T., {Berghmans}, D., {et~al.} 2024, arXiv
  e-prints, arXiv:2404.10577

\bibitem[{{R{\'e}ville} {et~al.}(2020){R{\'e}ville}, {Velli}, {Panasenco},
  {Tenerani}, {Shi}, {Badman}, {Bale}, {Kasper}, {Stevens}, {Korreck},
  {Bonnell}, {Case}, {de Wit}, {Goetz}, {Harvey}, {Larson}, {Livi},
  {Malaspina}, {MacDowall}, {Pulupa}, \& {Whittlesey}}]{reville2020}
{R{\'e}ville}, V., {Velli}, M., {Panasenco}, O., {et~al.} 2020, \apjs, 246, 24

\bibitem[{{Riley} {et~al.}(2019){Riley}, {Downs}, {Linker}, {Mikic},
  {Lionello}, \& {Caplan}}]{riley2019}
{Riley}, P., {Downs}, C., {Linker}, J.~A., {et~al.} 2019, \apjl, 874, L15

\bibitem[{{Sharma} \& {Morton}(2023)}]{sharma2023}
{Sharma}, R. \& {Morton}, R.~J. 2023, Nature Astronomy, 7, 1301

\bibitem[{{Shetye} {et~al.}(2021){Shetye}, {Verwichte}, {Stangalini}, \&
  {Doyle}}]{shetye2021}
{Shetye}, J., {Verwichte}, E., {Stangalini}, M., \& {Doyle}, J.~G. 2021, \apj,
  921, 30

\bibitem[{{Shi} {et~al.}(2021){Shi}, {Van Doorsselaere}, {Guo}, {Karampelas},
  {Li}, \& {Antolin}}]{shi2021}
{Shi}, M., {Van Doorsselaere}, T., {Guo}, M., {et~al.} 2021, \apj, 908, 233

\bibitem[{{Shoda} {et~al.}(2019){Shoda}, {Suzuki}, {Asgari-Targhi}, \&
  {Yokoyama}}]{shoda2019}
{Shoda}, M., {Suzuki}, T.~K., {Asgari-Targhi}, M., \& {Yokoyama}, T. 2019,
  \apjl, 880, L2

\bibitem[{{Shrivastav} {et~al.}(2023){Shrivastav}, {Pant}, {Berghmans},
  {Zhukov}, {Van Doorsselaere}, {Petrova}, {Banerjee}, {Lim}, \&
  {Verbeeck}}]{shrivastav2024}
{Shrivastav}, A.~K., {Pant}, V., {Berghmans}, D., {et~al.} 2023, arXiv
  e-prints, arXiv:2304.13554

\bibitem[{{Sishtla} {et~al.}(2022){Sishtla}, {Pomoell}, {Kilpua}, {Good},
  {Daei}, \& {Palmroth}}]{sishtla2022}
{Sishtla}, C.~P., {Pomoell}, J., {Kilpua}, E., {et~al.} 2022, \aap, 661, A58

\bibitem[{{Suzuki} \& {Inutsuka}(2005)}]{suzuki2005}
{Suzuki}, T.~K. \& {Inutsuka}, S.-i. 2005, \apjl, 632, L49

\bibitem[{{Terradas} {et~al.}(2008){Terradas}, {Andries}, {Goossens},
  {Arregui}, {Oliver}, \& {Ballester}}]{terradas2008b}
{Terradas}, J., {Andries}, J., {Goossens}, M., {et~al.} 2008, \apjl, 687, L115

\bibitem[{{Thurgood} {et~al.}(2014){Thurgood}, {Morton}, \&
  {McLaughlin}}]{thurgood2014}
{Thurgood}, J.~O., {Morton}, R.~J., \& {McLaughlin}, J.~A. 2014, \apjl, 790, L2

\bibitem[{{Tian} {et~al.}(2012){Tian}, {McIntosh}, {Wang}, {Ofman}, {De
  Pontieu}, {Innes}, \& {Peter}}]{tian2012}
{Tian}, H., {McIntosh}, S.~W., {Wang}, T., {et~al.} 2012, \apj, 759, 144

\bibitem[{{Tiwari} {et~al.}(2021){Tiwari}, {Morton}, \&
  {McLaughlin}}]{tiwari2021}
{Tiwari}, A.~K., {Morton}, R.~J., \& {McLaughlin}, J.~A. 2021, \apj, 919, 74

\bibitem[{{Tomczyk} \& {McIntosh}(2009)}]{tomczyk2009}
{Tomczyk}, S. \& {McIntosh}, S.~W. 2009, \apj, 697, 1384

\bibitem[{Tomczyk {et~al.}(2007)Tomczyk, McIntosh, Keil, Judge, Schad, Seeley,
  \& Edmondson}]{tomczyk2007}
Tomczyk, S., McIntosh, S.~W., Keil, S.~L., {et~al.} 2007, Science, 317, 1192

\bibitem[{{van der Holst} {et~al.}(2014){van der Holst}, {Sokolov}, {Meng},
  {Jin}, {Manchester}, {T{\'o}th}, \& {Gombosi}}]{vanderholst2014}
{van der Holst}, B., {Sokolov}, I.~V., {Meng}, X., {et~al.} 2014, \apj, 782, 81

\bibitem[{{Van Doorsselaere} {et~al.}(2014){Van Doorsselaere}, {Gijsen},
  {Andries}, \& {Verth}}]{vd2014}
{Van Doorsselaere}, T., {Gijsen}, S.~E., {Andries}, J., \& {Verth}, G. 2014,
  \apj, 795, 18

\bibitem[{{Van Doorsselaere} {et~al.}(2020{\natexlab{a}}){Van Doorsselaere},
  {Li}, {Goossens}, {Hnat}, \& {Magyar}}]{vd2020}
{Van Doorsselaere}, T., {Li}, B., {Goossens}, M., {Hnat}, B., \& {Magyar}, N.
  2020{\natexlab{a}}, \apj, 899, 100

\bibitem[{{Van Doorsselaere} {et~al.}(2024){Van Doorsselaere}, {Magyar},
  {Sieyra}, \& {Goossens}}]{vd2024}
{Van Doorsselaere}, T., {Magyar}, N., {Sieyra}, M.~V., \& {Goossens}, M. 2024,
  Journal of Plasma Physics, 90, 905900113

\bibitem[{{Van Doorsselaere} {et~al.}(2008){Van Doorsselaere}, {Nakariakov}, \&
  {Verwichte}}]{vd2008}
{Van Doorsselaere}, T., {Nakariakov}, V.~M., \& {Verwichte}, E. 2008, \apjl,
  676, L73

\bibitem[{{Van Doorsselaere} {et~al.}(2020{\natexlab{b}}){Van Doorsselaere},
  {Srivastava}, {Antolin}, {Magyar}, {Vasheghani Farahani}, {Tian}, {Kolotkov},
  {Ofman}, {Guo}, {Arregui}, {De Moortel}, \& {Pascoe}}]{vd2020ssrv}
{Van Doorsselaere}, T., {Srivastava}, A.~K., {Antolin}, P., {et~al.}
  2020{\natexlab{b}}, \ssr, 216, 140

\bibitem[{{Velli} {et~al.}(1989){Velli}, {Grappin}, \& {Mangeney}}]{velli1989}
{Velli}, M., {Grappin}, R., \& {Mangeney}, A. 1989, Physical Review Letters,
  63, 1807

\bibitem[{{Verdini} {et~al.}(2009){Verdini}, {Velli}, \&
  {Buchlin}}]{verdini2009}
{Verdini}, A., {Velli}, M., \& {Buchlin}, E. 2009, \apjl, 700, L39

\bibitem[{{Verdini} {et~al.}(2010){Verdini}, {Velli}, {Matthaeus}, {Oughton},
  \& {Dmitruk}}]{verdini2010}
{Verdini}, A., {Velli}, M., {Matthaeus}, W.~H., {Oughton}, S., \& {Dmitruk}, P.
  2010, \apjl, 708, L116

\bibitem[{{Wang} {et~al.}(2012){Wang}, {Ofman}, {Davila}, \& {Su}}]{wang2012}
{Wang}, T., {Ofman}, L., {Davila}, J.~M., \& {Su}, Y. 2012, \apjl, 751, L27

\bibitem[{{Zank} {et~al.}(2012){Zank}, {Dosch}, {Hunana}, {Florinski},
  {Matthaeus}, \& {Webb}}]{zank2012}
{Zank}, G.~P., {Dosch}, A., {Hunana}, P., {et~al.} 2012, \apj, 745, 35

\bibitem[{{Zank} {et~al.}(1996){Zank}, {Matthaeus}, \& {Smith}}]{zank1996}
{Zank}, G.~P., {Matthaeus}, W.~H., \& {Smith}, C.~W. 1996, \jgr, 101, 17093

\end{thebibliography}
%

\begin{appendix}

\section{Relation between wave energy $w$ and $Q$-density $W$}\label{sec:energy}
We have previously introduced the wave energy density in Eq.~\ref{eq:energy}, which we repeat here for convenience:
\begin{equation*}
	w^\pm = \frac{\rho_0 (\vec{\delta Z}^{\,\pm})^2}{4}.
\end{equation*}
We had also introduced the $Q$-density $W^\pm$ in Eq.~\ref{eq:qdensity}:
\begin{equation*}
    W^\pm = \frac{\rho_0}{4} \left(\vec{\delta Q}^\pm_\perp\right)^2.
\end{equation*}
These two expressions apparently do not coincide. In this section, we explore the interrelation between these two expressions.
To calculate the $Q$-density, we start from Eqs.~\ref{eq:vr}-\ref{eq:bz} to obtain
\begin{align}
    \delta Q_r^\pm & = \frac{\partial {\cal R}}{\partial r} \frac{\Omega \mp \alpha k_z B_0}{\rho_0(\Omega^2-\omega_\mathrm{A}^2)} \cos{\varphi}
\sin{(k_z z-\omega t)},\\
	\delta Q_\varphi^\pm & = -\frac{{\cal R}}{ r} \frac{\Omega \mp \alpha k_z B_0}{\rho_0(\Omega^2-\omega_\mathrm{A}^2)}
\sin{\varphi}
\sin{(k_z z-\omega t)}.
\end{align}
As discussed in \citet{vd2024}, we compute $\alpha$ in order to describe the wave in the co-moving waveframe. Thus, we have that
\begin{equation}
    \Omega = \mp \alpha k_zB_0.
\end{equation}
We thus find that (for example) the positive, downward wave energy is
\begin{equation}
    W^+=\frac{\Omega^2}{\rho_0(\Omega^2-\omega_\mathrm{A}^2)^2}\left(\left(\frac{\partial {\cal R}}{\partial r}\right)^2 \cos^2{\varphi} +
\left(\frac{{\cal R}}{ r}\right)^2\sin^2{\varphi}\right) \sin^2{(k_z z-\omega t)}
\end{equation}
and $W^-=0$. Indeed, the downward propagating wave is solely described  by $\vec{\delta Q}^+_\perp$. Now, we take the cross-sectional and wavelength
average of the $Q$-density $\langle W^\pm \rangle$,
where the radial integral is taken for the volume ($r\in [0,\gamma R]$) in which this plume is the dominant structure \citep{vd2014}. We relate this
outer boundary $\gamma R$ to the filling factor $f$ of the corona through the relation of \citet{vd2014}
\begin{equation}
    \gamma^2=\frac{1}{f}.
\end{equation}
Using a similar procedure
as \citet{vd2014} and \citet{vd2024}, we have as cross-sectionally averaged $Q$-density $\langle W^\pm \rangle$ (see Eq.~\ref{eq:cross-average}):
\begin{align}
    \langle W^\pm \rangle &= \pi A^2\Omega^2 \int_0^R \frac{rdr}{R^2\rho_\mathrm{i}(\Omega^2-\omega_\mathrm{Ai}^2)^2} + \int_R^{\gamma R}
\frac{R^2rdr}{r^4\rho_\mathrm{e}(\Omega^2-\omega_\mathrm{Ae}^2)^2},\\
	&= \frac{\pi A^2\Omega^2}{2} \frac{1}{\rho_\mathrm{e}^2(\Omega^2-\omega_\mathrm{Ae}^2)^2}(\rho_\mathrm{i}+\rho_\mathrm{e}(1-f)),\\
	& = \pi R^2 \Upsilon^2 \frac{\rho_\mathrm{i}+\rho_\mathrm{e}(1-f)}{2}. \label{eq:qdens}
\end{align}
In the second line, we have used the dispersion relation for the kink wave with
$\rho_\mathrm{i}(\Omega^2-\omega_\mathrm{Ai}^2)=\rho_\mathrm{e}(\Omega^2-\omega_\mathrm{Ae}^2)$, and in the third line we have used the expression
for the wave amplitude $\Upsilon$ (see Eq.~\ref{eq:amplitudes}). The term without $f$ corresponds exactly to the expression of the energy density
$\langle w\rangle$ \citep[see e.g. Eq.~10 in][]{vd2014}. There is only a difference in the term proportional to the filling factor $f$. Note that if
there are no other structures in the corona, we would integrate until $r\to \infty$ and consequently $f=0$, and hence the expression of the energy
density $\langle w\rangle$ is in that case the same as $\langle W^\pm \rangle$.

To explain this (seemingly remarkable) coincidence between the expressions for $\langle W^\pm\rangle$ and $\langle w\rangle$, we start again from the
expression for the wave energy density $w$ (see Eq.~\ref{eq:energy}) and the $Q$-density (Eq.~\ref{eq:qdensity}).
\begin{align}
    W^\pm &= \frac{\rho_0}{4} (\vec{\delta V}_\perp \pm \alpha \vec{\delta B}_\perp )^2 = \frac{\rho_0}{4} (\delta V_\perp^2 + \alpha^2\delta
B_\perp^2\pm 2 \alpha \vec{\delta B}_\perp\cdot \vec{\delta V}_\perp),\\
	w &= w^++w^-=\frac{\rho_0}{2} ( \delta V_\perp^2 + \frac{\delta
B_\perp^2}{\mu\rho_0}).
\end{align}
The equations are seemingly different, but this difference is only apparent indeed. By construction, we have that
\begin{equation}
    \vec{\delta Q}^\mp_\perp=0
\end{equation}
for the oppositely propagating wave. Thus, we must have
\begin{equation}
    \vec{\delta V}_\perp \mp \alpha \vec{\delta B}_\perp = 0, \quad \mbox{or} \quad \vec{\delta V}_\perp = \pm \alpha \vec{\delta B}_\perp.
\end{equation}
Then, we see that the $Q$-density equals
\begin{equation}
    W^\pm = \rho_0 \delta V^2_\perp.
\end{equation}
This is equal to the wave energy $w$ in case of equipartition of the wave (over the domain in which the wave energy is averaged). Thus, for
well-behaved waves \citep[such as the kink wave, which has equipartition,][]{goossens2013}, the $Q$-density is exactly equal to the wave energy
density. As explained in \citet{goossens2013}, the equipartition between magnetic and kinetic energy for kink waves is only satisfied when
considering the entire domain. Here however, the spatial domain is limited to a distance $\gamma R$, so that the equipartition is not exact,
explicited by the presence of the term containing the filling factor $f$.
\end{appendix}

\end{document}